\documentclass[12pt]{article}
\begin{document}

\title{Some Limit Theorems for Linear
Oscillators with Noise in the Coefficients\footnote{extended version
including proofs of a contribution by V. B. at the 
Workshop "Nonlinear and Stochastic Beam Dynamics in Accelerators -
 a Challenge to Theoretical and Computational Physics" 
L\"{u}neburg (1997)}}

\author{V. Balandin$^{*}$ and $\;$ H. Mais$^{\dagger}$\\
\\
$^{*}\;$\small\it Institute for Nuclear Research of RAS, \\
\small\it 60th October Anniversary Pr., 7a, Moscow 117 312, Russia\\
$^{\dagger}\;$\small\it Deutsches Elektronen-Synchrotron DESY, Hamburg
}
\date{\today}
\maketitle

\begin{abstract}
Using the tools and methods developed in 
\cite{CandDis} limit theorems are proven for 
the linear oscillator with random coefficients.
The asymptotic behaviour of the moments is
studied in detail. The technique 
 presented in this paper  can be
applied to general linear systems with noise and is well
suited for the investigation of stochastic beam dynamics
in accelerators.
\end{abstract}

\clearpage
\tableofcontents
\clearpage

\section{Linear Oscillator with Noise in the Coefficients}

Starting point of our investigation is a 
nondegenerate ($\;\omega_0 \neq 0\;$)
damped linear oscillator
under the influence of noise
\begin{eqnarray}
\ddot{x} \:+\: \varepsilon \:
(\gamma(t) \:+\: \varepsilon \: \alpha) \: \dot{x} \:+\:
\omega_0^2 \:
\left(1 \:+\: \frac{\varepsilon}{\omega_0} \: \eta(t)\right)
\:x \;=\; 
\varepsilon \: \omega_0 \: \xi(t)
\label{m01}
\end{eqnarray}
or written as a 
system of two first-order 
differential equations
\begin{eqnarray}
\left\{
\begin{array}{l}
\dot{x} \;=\; \hspace{0.3cm} \omega_0 \: z \\
\\
\dot{z} \;=\; -\omega_0 \: x \:+\: 
\varepsilon \:
\left(\xi \:-\: \gamma \: z \:-\: \eta \: x \right) \:-\:
\varepsilon^2 \:\alpha\: z
\end{array}
\right.
\label{m02}
\end{eqnarray}

$\:\varepsilon\:$ is small parameter 
$\;|\varepsilon|\:<\:1\;$. 

The $\varepsilon^2$ proportionality of the deterministic
term in the damping part is connected
with the fact that we will discuss the dynamics on time scales 
$O(1/\varepsilon^2)$ (it is the minimum time scale where the
stochastic effects could essentially influence the dynamics of 
our oscillator). If the damping will be weaker it will
not affect the dynamics and we can neglect it, 
and if it will be stronger it will
completely change the picture of the dynamics, the typical time
scales become exponentially large 
$O(\exp(1/\varepsilon^a)), a > 0$
for positive damping 
and it will require other methods
(see, for example \cite{Wentz}) that are beyond the scope of this
paper.

Noise has been introduced in the damping part
($\gamma(t)$), as a modulation of the 
frequency $\omega_0$ ($\eta(t)$) and as an
external driving force $\xi(t)$.

As a model of noise we shall take stochastic
processes defined by the following scalar products
\begin{eqnarray}
\eta(t)   = \vec b(t) \cdot \vec y(t), \hspace{0.6cm}
\xi(t)    = \vec h(t) \cdot \vec y(t), \hspace{0.6cm}
\gamma(t) = \vec d(t) \cdot \vec y(t)
\nonumber
\end{eqnarray}
with nonrandom $n$-dimensional 
vectors $\;\vec{b}$, $\;\vec{h}\;$ and $\;\vec{d}\;$ 
which are quasiperiodic in $\;t\;$ 
and which can be 
expanded into Fourier series
\begin{eqnarray}
\vec{b}(t) \;=\; 
\sum \limits_{m = -\infty}^{+\infty}
\vec{b}_{m} \exp(i \nu_m t),
\hspace{1.0cm} 
\vec{b}_{-m} = \left(\vec{b}_{m}\right)^{*}
\nonumber
\end{eqnarray}
\begin{eqnarray}
\vec{h}(t) \;=\; 
\sum \limits_{m = -\infty}^{+\infty}
\vec{h}_{m} \exp(i \nu_m t),
\hspace{1.0cm}
\vec{h}_{-m} = \left(\vec{h}_{m}\right)^{*}
\nonumber
\end{eqnarray}
\begin{eqnarray}
\vec{d}(t) \;=\; 
\sum \limits_{m = -\infty}^{+\infty}
\vec{d}_{m} \exp(i \nu_m t),
\hspace{1.0cm}
\vec{d}_{-m} = \left(\vec{d}_{m}\right)^{*} 
\nonumber
\end{eqnarray}
with real frequencies $\;\nu_m\;$ satisfying the condition 
\begin{eqnarray}
\nu_{l}\:+\:\nu_{m}\:=\:0 \;\Leftrightarrow\; 
m\:+\:l\:=\:0.
\nonumber
\end{eqnarray}
In the main part of this paper
the vector $\;\vec y(t) \in R^n\;$ is assumed to be a 
solution of the linear system of Ito's stochastic 
differential equations
\begin{eqnarray}
d \, \vec{y} \;=\; A \, \vec{y} \cdot dt \;+\; B\, d \vec{w}(t)
\label{ss1}
\end{eqnarray}
where $\;A\;$ and $\;B\;$ are ($n \times n$) 
and ($n \times r$) real constant matrices respectively, 
and $\;\vec{w}(t)\;$ 
is an $r$-dimensional Brownian motion,
other choices for the noise model will be
described later on.

As smoothness properties of the vector functions
$\;\vec b$, $\;\vec h\;$ and $\;\vec d\;$ we shall require 
the convergence of the series
\footnote{
For a complex vector $\vec w \in C^n$ we use the usual 
spherical norm 
$\;|\vec w| = \sqrt{\vec w \cdot \vec w} = 
\sqrt{w_1 \cdot w_1^* + \ldots +w_n \cdot w_n^* }\;$
and for a ($n \times n$) matrices with complex 
coefficients we shall use the norm  $\;|M| = \sqrt{\lambda}\;$  
where $\lambda$ is the greatest eigenvalue of the matrix 
$M^* M$, which is compatible with the spherical norm for vectors.}
\begin{eqnarray}
\sum \limits_{m = -\infty}^{+\infty}
|\nu_m|^p 
\left(
\left|\vec{b}_{m}\right| +
\left|\vec{h}_{m}\right| +
\left|\vec{d}_{m}\right|\right) \;<\;\infty,
\hspace{0.7cm}
p\:=\:0,1
\label{cond1}
\end{eqnarray}

We denote
\begin{eqnarray}
F \;=\;
\left\{\:p\:\in\: Z\; : \;
|\vec{b}_p|+|\vec{h}_p|+|\vec{d}_p|
\:\neq \: 0
\right\}
\nonumber
\end{eqnarray}
and introduce
\begin{eqnarray}
F_k \;=\;
\left\{(p,\:q)\:\in\: F \times F\; : \;
|\nu_p + \nu_q - k \:\omega_0|
\:\neq \: 0
\right\}.
\nonumber
\end{eqnarray}

Besides the smoothness condition (\ref{cond1}) we also
require
\begin{eqnarray}
\min_{k\:\in\:\{0,1,2,3,4\}}
\hspace*{0.3cm}
\inf_{(p,\:q)\:\in\: F_k} |\nu_p \:+\: \nu_q \:-\: k \:\omega_0|
\;\geq\;\delta_f^2\;>\;0.
\label{cond2}
\end{eqnarray}
The condition (\ref{cond2}) does not exclude resonances
but requires them to be isolated.
This  can be easily changed to some kind of
Diophantine conditions with increasing smoothness properties
(\ref{cond1}).
Note that (\ref{cond2}) is always satisfied for periodic
functions (i.e. $\nu_p \:=\: p \cdot \nu$) and for finite
trigonometrical polynomials with arbitrary frequencies.

In this paper we will assume that all eigenvalues 
$\;\lambda_k\;$ of the matrix $\;A\;$ 
in (\ref{ss1})
have negative real parts, 
i.e.
\begin{eqnarray}
Re \:\lambda_k \;\leq \;-\delta_s^2 \;<\; 0,
\hspace{0.7cm} k = 1, \ldots, n
\label{SS11}
\end{eqnarray}

From this it follows 
(see, for example \cite{Karatzas})
that if the initial random vector 
$\;\vec{y}_0\;$, 
independent of the $r$-dimensional
Brownian motion 
$\;\vec{w}(t)-\vec{w}(0)\;$ for $\;0 < t < \infty\;$,
has a normal distribution with mean value
$\;\left< \vec{y}_0 \right> \;=\; \vec{0}\;$ and covariance
matrix
\begin{eqnarray}
\left<\vec{y}_0 \cdot \vec{y}_0^{\top}\right> \;= \;
\int \limits_{0}^{\infty} 
\exp(\tau A) \:B B^{\top} \exp(\tau A^{\top}) \:d \tau\;
\stackrel{\rm def}{=}\; D
\nonumber
\end{eqnarray}
then the solution 
of (\ref{ss1})
$\;\vec{y}(t, \vec{y}_0)\;$ is a stationary,
zero-mean Gaussian process, with covariance function
\begin{eqnarray}
\rho (\tau) \;=\;
\left\{
\begin{array}{lll}
\exp(\tau A) \: D &; & \tau \geq 0\\
\\
D \: \exp(\tau A^{\top}) &; & \tau \leq 0
\end{array}
\right.
\label{sden}
\end{eqnarray}

Although,
later on we shall not restrict the initial conditions for
$\;\vec{y}\;$ in our noise model to be equal 
to the above mentioned initial conditions generating 
stationary solutions of the system (\ref{ss1})
\footnote{For simplicity we even shall take the initial condition
to be a point in $n$-dimensional Euclidean space, but if one will
follow the proofs of the theorems
it will be clear that all results of this paper
will be correct if we  use as initial condition an arbitrary
random vector, 
independent of the $r$-dimensional
Brownian motion $\vec{w}(t)-\vec{w}(0)$ for $0 < t < \infty$,
additionally assuming that
some moments of $\vec{y}_0$ are finite.},
all results will nevertheless be expressed in terms of
the spectral density
associated with the covariance function (\ref{sden}) 
$\;\Psi(\omega)\:=\:\Psi_c(\omega)\:-\:i\:\Psi_s(\omega)\;$
where
\footnote{Note that if the matrices $A$ and $B^{-1}$
commute we use notation $\frac{A}{B}$ for the product $A B^{-1}$.}
\begin{eqnarray}
\Psi_c(\omega) \;= \;
\int \limits_{0}^{\infty} 
\cos(\omega \tau)\rho(\tau) d \tau
\;=\; - \frac{A}{A^2 + \omega^2 I} \cdot D
\nonumber
\end{eqnarray}
\begin{eqnarray}
\Psi_s(\omega) \;= \;
\int \limits_{0}^{\infty} 
\sin(\omega \tau)\rho(\tau) d \tau
\;=\;  \hspace{0.3cm}
\frac{\omega I}{A^2 + \omega^2 I} \cdot D
\nonumber
\end{eqnarray}
For further purposes
let us note that independently from real $\:\omega\:$ 
the norm of the matrix $\;\Psi(\omega)\;$ admits the
estimate
\begin{eqnarray}
\left|\Psi(\omega)\right| \;\leq \; \bar{C}
\label{cond3}
\end{eqnarray}
where $\,\bar{C}\,$ is some positive constant whose exact
value depends on $\;\delta_s^2\;$ and 
$\;\left|B B^{\top}\right|\;$ and  
is unimportant for us.

\section{Special Basis in the Space of Polynomials}

Often, the influence of noise in systems such as 
(\ref{m01}) is studied by considering its
influence on the unperturbed invariants of motion such
as energy
\begin{eqnarray}
r = \frac{1}{2}\left(x^2 + z^2 \right)
\nonumber
\end{eqnarray}
or functions of the energy. For our later study
of arbitrary moments we introduce a special time
dependent (non-autonomous) 
basis in the space of polynomials.

For all nonnegative integers $m$, $k$ we define
\begin{eqnarray}
I_{m,\:k} \: = \:
\exp\left(i \: (m - k) \: \omega_0 \: t \right)
\left(\frac{x + i z}{2}\right)^m
\left(\frac{x - i z}{2}\right)^k \; 
\nonumber
\end{eqnarray}
It is easy to check that the functions
introduced above admit the following properties 
\begin{eqnarray}
\begin{array}{ll}
{\bf a.} & 
\left( 
\frac{\partial}{\partial t}\;+\;
\omega_0 \: 
\left(
z \: \frac{\partial}{\partial x} - 
x \: \frac{\partial}{\partial z}
\right)
\right)
\: I_{m, \: k} \: = \: 0 \\
\\
{\bf b.} & I_{m_1, \: k_1} \cdot I_{m_2, \: k_2} 
\: = \: I_{m_1 + m_2, \; k_1 + k_2} \\
\\
{\bf c.} & I_{m, \: k} \: = \: I_{k, \: m}^* \\
\\
{\bf d.} & I_{m, \: m} \: = \: \left(\frac{r}{2}\right)^m \\
\\
{\bf e.} & |I_{m, \: k}|^2 \: = \: 
I_{m, \: k} \: \cdot \: I_{m, \: k}^* 
\: = \:
\left(\frac{r}{2}\right)^{m + k}\\
\end{array}
\nonumber
\end{eqnarray}
Representing $\;x\;$ and $\;z\;$ as
\begin{eqnarray}
x \;=\; \frac{x + iz}{2}\:+\: \frac{x - iz}{2} \;=\;
\hspace*{0.3cm}
\exp(i\omega_0 t) I_{0,\:1} +
\hspace*{0.3cm}
\exp(-i\omega_0 t) I_{1,\:0}
\nonumber
\end{eqnarray}
\begin{eqnarray}
z \;=\; \frac{x + iz}{2i}\:-\: \frac{x - iz}{2i} \;=\;
i\exp(i\omega_0 t) I_{0,\:1} -
i\exp(-i\omega_0 t) I_{1,\:0}
\nonumber
\end{eqnarray}
and using property {\bf b}
we can express $\;x^{m-k} \cdot z^k\;$ ($0 \leq k \leq m$) 
with the help of the binomial theorem
in the form of a linear combination of the functions $\;I_{p,\;q}$
\begin{eqnarray}
x^{m-k} \cdot z^k \;=\; 
(i)^k \cdot
\nonumber
\end{eqnarray}
\begin{eqnarray}
\cdot
\sum \limits_{p\:=\:0}^{m-k}
\sum \limits_{q\:=\:0}^{k}
(-1)^q
\left(
\begin{array}{c}
m-k\\
p
\end{array}
\right)
\left(
\begin{array}{c}
k\\
q
\end{array}
\right)
\exp(i(m-2(p+q))\:\omega_0 t)\cdot
I_{p+q,\:m-(p+q)}.
\nonumber
\end{eqnarray}

For $\;m \neq k\;$ $\;I_{m,\:k}\;$ are functions with 
complex values. However, we can also use as basis real 
valued functions 
$\;U_{m,\:k}\;$ and $\;V_{m,\:k}\;$ which are defined by
\begin{eqnarray}
U_{m,\:k} \;=\; \frac{I_{m,\:k} \:+\:I_{k,\:m}}{2}
\;=\;U_{k,\:m},
\hspace{1.0cm}
V_{m,\:k} \;=\; \frac{I_{m,\:k} \:-\:I_{k,\:m}}{2i}
\;=\;-V_{k,\:m}.
\nonumber
\end{eqnarray}

Note further that the functions $\;U_{m,\:k}\;$ and
$\;V_{m,\:k}\;$ 
can be easily expressed
through the real valued functions 
$\;\bar{U}_{m,\:k}\;$ and $\;\bar{V}_{m,\:k}\;$
\begin{eqnarray}
\bar{U}_{m,\:k} \;=\; 
\frac{
\left(\frac{x + i z}{2}\right)^m
\left(\frac{x - i z}{2}\right)^k
\;+\:
\left(\frac{x + i z}{2}\right)^k
\left(\frac{x - i z}{2}\right)^m
}{2}
\nonumber
\end{eqnarray}
\begin{eqnarray}
\bar{V}_{m,\:k} \;=\; 
\frac{
\left(\frac{x + i z}{2}\right)^m
\left(\frac{x - i z}{2}\right)^k
\;-\:
\left(\frac{x + i z}{2}\right)^k
\left(\frac{x - i z}{2}\right)^m
}{2\,i}
\nonumber
\end{eqnarray}
which do not depend on time $t$
with help of the following simple formula
\begin{eqnarray}
\left(
\begin{array}{c}
U_{m,\:k}\\
\\
V_{m,\:k}
\end{array}
\right)
=
\left(
\begin{array}{rr}
 \cos((m-k)\:\omega_0 t) & -\sin((m-k)\:\omega_0 t) \\
\\
 \sin((m-k)\:\omega_0 t) &  \cos((m-k)\:\omega_0 t) 
\end{array}
\right)
\cdot
\left(
\begin{array}{c}
\bar{U}_{m,\:k}\\
\\
\bar{V}_{m,\:k}
\end{array}
\right)
\nonumber
\end{eqnarray}

\section{Stopped Process}

\hspace*{0.5cm}
Although a suitable choice for $A$ and $B$ in (\ref{ss1})
allows one to approximate a wide range of spectral functions
(with appropriate choice of $A$ and $B$ one can obtain for
the $y_1$ component of the vector $\vec{y}$ 
every spectral function which is the ratio of two 
polynomials),
the solution of this equation has the disadvantage that it
also allows with positive probability arbitrary big excursions
during finite fixed time intervals. 
In order to remove this effect and also to apply our 
proof technique  we
have to freeze and truncate the process.

Let $\:c(\varepsilon)\:$ be some positive function of 
$\:\varepsilon\:$ defined on the set $\:\varepsilon \,\neq\, 0\,$.
For every natural $\:m\:$ and for every point
$\;\vec{y}_0\: \in\: R^n\;$ we introduce a random value
\begin{eqnarray}
\tau_m^{\varepsilon} \;=\; \tau_m^{\varepsilon}(\vec{y}_0)
\;=\; \inf 
\left\{
t \geq 0: \;\; (t, \:\vec{y}(t)) \:\not\in\:
\left[0, \: m \right) \times 
\left\{
\vec{y} : \;\; |\vec{y}| \:<\: c(\varepsilon)
\right\}
\right\} 
\nonumber
\end{eqnarray}
where $\;\vec{y}(t)\;$ is the solution of the system (\ref{ss1})
which with probability one satisfies the initial condition
$\;\vec{y}(0)\: =\: \vec{y}_0\,$.
So with probability one for $\;m_1 \:\leq\: m_2\;$
\begin{eqnarray}
0 \;\leq\; \tau_{m_1}^{\varepsilon} \;\leq\; 
\tau_{m_2}^{\varepsilon}. 
\nonumber
\end{eqnarray}
Then with probability one there exists a limit (finite or infinite) 
when $\:m \rightarrow \infty\:$ of the sequence 
$\;\tau_{m}^{\varepsilon}\;$ which we will denote as
\begin{eqnarray}
\tau^{\varepsilon}(\vec{y}_0) \;\stackrel{{\rm def}}{=}\; 
\lim_{m \rightarrow \infty} \tau_{m}^{\varepsilon}(\vec{y}_0).
\nonumber
\end{eqnarray}
In other words
$\;\tau^{\varepsilon}(\vec{y}_0)\;$ is the exit time from an open
ball $\;|\vec{y}| \:<\: c(\varepsilon)\;$ for the solution of (\ref{ss1})
starting with probability one from initial point $\:\vec{y}_0$.
Note that if the matrix $\;B B^{\top}\;$ is nondegenerate then
this exit time is finite with probability one. 

The joint solution of the systems (\ref{m02}), (\ref{ss1})
$\;(x(t),\,z(t),\, \vec{y}(t))\;$
is a Markovian diffusion process in ($n+2$)-dimensional
Euclidean space. 
Let $\:s_{t}^{\varepsilon} \:=\: 
\min\left\{t, \:\tau^{\varepsilon}\right\}\:$.
For the noise model (\ref{ss1}) 
for reasons which we explained above
we shall not
study the moments
of the stochastic process 
$\;(x(t),\,z(t),\, \vec{y}(t))\,$, 
but the moments of
the stochastic process 
$\;(x(s_{t}^{\varepsilon}),\,z(s_{t}^{\varepsilon}),\, 
\vec{y}(s_{t}^{\varepsilon}))\,$ 
(stopped process). We shall use the time scale
$\;O\left(\varepsilon^{-2}\right)\;$ and the difference
between $\:t\:$ and $\:s_{t}^{\varepsilon}\:$ for
this time scale can be estimated with the help of
the following  

{\bf Theorem A:$\;$} 
{\sl
There exist positive constants $\:a\:$ and $\:b\:$ 
so that for any initial point $\:\vec{y}_0\:$ and for any 
positive $\:L\:$
\begin{eqnarray}
P \left( \tau^{\varepsilon} < 
\frac{L}{\varepsilon^2}\right) \;\leq\; 
\left(\exp(a \,|\vec{y}_0|^2) \:+\: a \,\frac{L}{\varepsilon^2}\right) 
\exp(-b\, c^2(\varepsilon)) 
\label{teA}
\end{eqnarray}
}

Rewriting the left hand side of the inequality (\ref{teA})
in the form
\begin{eqnarray}
P \left( \tau^{\varepsilon} < 
\frac{L}{\varepsilon^2}\right) \;=\;
P \left(
\max_{0 \leq t \leq L / \varepsilon^2}
\left|\: 
t\: -\:s_t^{\varepsilon}\:
\right|\:>\:0
\right)
\nonumber
\end{eqnarray}
we see that on the time scale considered 
the measure of the set where
$\;t\: \neq \:s_t^{\varepsilon}\;$ 
will go to zero as 
$\;\varepsilon\:\rightarrow\:0\;$
if 
$\;c^2(\varepsilon)\:\rightarrow\:\infty\;$
faster then $\;b^{-1}\,\log\left(\varepsilon^{-2}\right)$.
On the other hand to apply the technique of our proof
we  require that
\begin{eqnarray}
\lim_{\varepsilon \rightarrow 0} \varepsilon c^3(\varepsilon) = 0
\nonumber
\end{eqnarray}
so that we can not allow $\:c(\varepsilon)\:$ go to infinity
too fast.

\section{Asymptotic Behaviour of Moments}

Let us introduce functions $\;\bar{c}_l(m,\:k)\;$ 
of integer arguments $\:m,\,k\:\geq\:0\:$
with the help of 
\begin{eqnarray} 
\bar{c}_1(m, \:k) \;=\;
\frac{m}{4}
\left(
\sum \limits_{\nu_l-\nu_p\:=\:\omega_0}
\left\{
\rule[0.2cm]{0cm}{0.2cm}
(m-1) \: \Psi^*(\omega_0 + \nu_p)\: 
\vec{h}_p \cdot
\left(\vec{b}_l + i \vec{d}_l\right) 
\;-\;
\right.
\right.
\nonumber
\end{eqnarray}
\begin{eqnarray} 
-\;(k+1)\:\Psi^*(2\omega_0 + \nu_p)\: 
\left(\vec{b}_p + i \vec{d}_p\right) 
\cdot 
\vec{h}_l
\;-
\nonumber
\end{eqnarray}
\begin{eqnarray} 
\left.
-\;k\: \Psi^*(\nu_p)\: 
\left(\vec{b}_p + i \vec{d}_p\right) 
\cdot 
\vec{h}_l \;-\;
k \:\Psi^*(\omega_0+\nu_p)\: 
\vec{h}_p \cdot
\left(\vec{b}_l - i \vec{d}_l\right) 
\rule[0.2cm]{0cm}{0.2cm}
\right\} \;+
\nonumber
\end{eqnarray}
\begin{eqnarray} 
\left.
+
\sum \limits_{\nu_l-\nu_p\:=\:-\omega_0}
\left\{
\rule[0.2cm]{0cm}{0.2cm}
m \Psi^{\top}(\nu_p)\: 
\left(\vec{b}_p^{\;*} - i \vec{d}_p^{\;*}\right) 
\cdot \vec{h}_l^{\;*} -
k \Psi^{\top}(\omega_0+\nu_p)\: 
\vec{h}_p^{\;*} \cdot
\left(\vec{b}_l^{\;*} - i \vec{d}_l^{\;*}\right) 
\rule[0.2cm]{0cm}{0.2cm}
\right\} 
\right)
\nonumber
\end{eqnarray}
\begin{eqnarray} 
\bar{c}_2(m, \:k) \;=\;
-\frac{m(m-1)}{4} 
\sum \limits_{\nu_l-\nu_p\:=\:2\omega_0}
\Psi^*(\omega_0 + \nu_p)\: 
\vec{h}_p \cdot \vec{h}_l
\nonumber
\end{eqnarray}
\begin{eqnarray} 
\bar{c}_3(m, \:k) \;=\;
\frac{m(m-1)}{4} \sum \limits_{\nu_l-\nu_p\:=\:3\omega_0}
\left\{
\rule[0.2cm]{0cm}{0.2cm}
\Psi^*(\omega_0 + \nu_p)\: 
\vec{h}_p \cdot
\left(\vec{b}_l - i \vec{d}_l\right) 
\;+\;
\right.
\nonumber
\end{eqnarray}
\begin{eqnarray} 
\left.
+\;\Psi^*(2\omega_0 + \nu_p)\: 
\left(\vec{b}_p + i \vec{d}_p\right) 
\cdot \vec{h}_l
\rule[0.2cm]{0cm}{0.2cm}
\right\}
\nonumber
\end{eqnarray}
\begin{eqnarray} 
\bar{c}_4(m, \:k) \;=\;
-\frac{m(m-1)}{4} 
\sum \limits_{\nu_l-\nu_p\:=\:4\omega_0}
\Psi^*(2\omega_0 + \nu_p)\: 
\left(\vec{b}_p + i \vec{d}_p\right) 
\cdot \left(\vec{b}_l - i \vec{d}_l\right)
\nonumber
\end{eqnarray}
\begin{eqnarray} 
\bar{c}_5(m, \:k) \;=\;
\frac{m}{4}
\left(
\sum \limits_{\nu_l-\nu_p\:=\:2\omega_0}
\left\{
\rule[0.2cm]{0cm}{0.2cm}
k \:\Psi^*(\nu_p)\: 
\left(\vec{b}_p + i \vec{d}_p\right) 
\cdot
\left(\vec{b}_l - i \vec{d}_l\right) 
\;+\;
\right.
\right.
\nonumber
\end{eqnarray}
\begin{eqnarray} 
+\;(k+1)\:\Psi^*(2\omega_0 + \nu_p)\: 
\left(\vec{b}_p + i \vec{d}_p\right) 
\cdot 
\left(\vec{b}_l - i \vec{d}_l\right) 
\;-
\nonumber
\end{eqnarray}
\begin{eqnarray} 
\left.
-\;(m-1)\:\Psi^*(2\omega_0 + \nu_p)\: 
\left(\vec{b}_p + i \vec{d}_p\right) 
\cdot 
\left(\vec{b}_l + i \vec{d}_l\right) 
\rule[0.2cm]{0cm}{0.2cm}
\right\} \;-
\nonumber
\end{eqnarray}
\begin{eqnarray} 
\left.
-\;\sum \limits_{\nu_l-\nu_p\:=\:-2\omega_0}
m \:\Psi^{\top}(\nu_p)\: 
\left(\vec{b}_p^{\;*} - i \vec{d}_p^{\;*}\right) 
\cdot
\left(\vec{b}_l^{\;*} - i \vec{d}_l^{\;*}\right) 
\right)
\nonumber
\end{eqnarray}
\begin{eqnarray} 
\bar{c}_6(m, \:k) \;=\;
\frac{m\:k}{4} \sum \limits_{p = -\infty}^{\infty}
\;\left[\:
\rule[0.2cm]{0cm}{0.2cm}
\Psi(\omega_0 + \nu_p) \; +\;
\Psi^*(\omega_0 + \nu_p) 
\:\right]\:
\vec{h}_p \cdot \vec{h}_p\;
\nonumber
\end{eqnarray}
\begin{eqnarray} 
\bar{c}_7(m, \:k) \;=\;
\nonumber
\end{eqnarray}
\begin{eqnarray} 
\hspace*{0.5cm}
- \:\frac{m + k}{2}\; \alpha \;+\;
\sum \limits_{p = -\infty}^{\infty}\;
\left\{
\frac{m\:k}{4} \:
\left[\:
\rule[0.2cm]{0cm}{0.2cm}
\Psi(\nu_p)\:+\:\Psi^*(\nu_p)\:\right]\:
\left(\vec{b}_p + i \vec{d}_p\right) 
\cdot \left(\vec{b}_p + i \vec{d}_p\right) \;-\; 
\right.
\nonumber
\end{eqnarray}
\begin{eqnarray} 
\hspace*{0.5cm}
-\:
\frac{m^2}{4}
\Psi(\nu_p)\:
\left(\vec{b}_p - i \vec{d}_p\right) 
\cdot \left(\vec{b}_p + i \vec{d}_p\right) \;-\; 
\frac{k^2}{4} 
\Psi^*(\nu_p)\:
\left(\vec{b}_p + i \vec{d}_p\right) 
\cdot \left(\vec{b}_p - i \vec{d}_p\right) \;+\; 
\nonumber
\end{eqnarray}
\begin{eqnarray} 
+
\left.
\left[\frac{(m+1)k}{4}
\Psi(2\omega_0+\nu_p)\:+\:
\frac{m(k+1)}{4} 
\Psi^*(2\omega_0+\nu_p)
\right]
\left(\vec{b}_p + i \vec{d}_p\right) 
\cdot \left(\vec{b}_p + i \vec{d}_p\right)
\right\}
\nonumber
\end{eqnarray}
By using (\ref{cond1}) and (\ref{cond3})
it is not hard to show that $\;c_l(m,\:k)\;$ 
are correctly defined
because the series converge absolutely
for every fixed values of $\,m\,$ and $\,k\,$. 

Now in  correspondence with 
an arbitrary
two index array $\:a_{m,\:k}\:$
and nonnegative integer $\:N\:$
\begin{eqnarray} 
m,\,k\;\geq\;0,
\hspace*{1.0cm}
m \:+\: k\;\leq\;N
\nonumber
\end{eqnarray}
we define a vector 
$\:\vec{\cal V}(a_{m,\:k};\,N)\:$ with
$\:(N+1)(N+2)/2\:$ components with the help of the rule
\begin{eqnarray} 
{\cal V}_{l}(a_{m,\:k};\,N) \;=\; a_{m,\:k},
\hspace*{1.0cm} 
l \;=\; \frac{(m+k)(m+k+1)}{2} \:+\:k\:+\:1
\nonumber
\end{eqnarray}
This ordering corresponds to the following ordering of the
elements of the array $a_{m, \:k}$ (take by rows)
\begin{eqnarray} 
\begin{array}{lllll}
a_{0,\:0} \\
a_{1,\:0} & a_{0,\:1} \\
a_{2,\:0} & a_{1,\:1} & a_{0,\:2} \\
\vdots \\
a_{N,\:0} & a_{N-1,\:1} & a_{N-2,\:2} &\ldots & a_{0,\:N}
\end{array}
\nonumber
\end{eqnarray}

Consider now the system of ordinary differential
equations with constant coefficients
\begin{eqnarray}
\frac{d}{d \tau}\:\vec{\cal V}(a_{m,\:k};\,N)\;=\;
\bar{\cal K}_{N} \; \vec{\cal V}(a_{m,\:k};\,N)
\label{srk}
\end{eqnarray}
generated with the help of the rule
\begin{eqnarray}
\begin{array}{llllll}
\frac{d}{d \tau}\: a_{m,\:k}& =
&
\bar{c}_2(m,\:k) \: a_{m-2,\:k} & + &
\bar{c}_2^{\;*}(k,\:m) \: a_{m,\:k-2} & + \\
\\
& & 
\bar{c}_1(m,\:k) \: a_{m-1,\:k} & + &
\bar{c}_1^{\;*}(k,\:m) \: a_{m,\:k-1} & + \\
\\
& & 
\bar{c}_3(m,\:k) \: a_{m-2,\:k+1} & + &
\bar{c}_3^{\;*}(k,\:m) \: a_{m+1,\:k-2} & + \\
\\
& &
\bar{c}_4(m,\:k) \: a_{m-2,\:k+2} & + &
\bar{c}_4^{\;*}(k,\:m) \: a_{m+2,\:k-2} & + \\
\\
& & 
\bar{c}_5(m,\:k) \: a_{m-1,\:k+1}  & + &
\bar{c}_5^{\;*}(k,\:m) \: a_{m+1,\:k-1} & + \\
\\
& & 
\bar{c}_6(m,\:k) \: a_{m-1,\:k-1}  & + &
\bar{c}_7(m,\:k) \: a_{m,\:k}
\end{array}
\label{rule_r}
\end{eqnarray}
where on the right hand side of (\ref{rule_r}) we take
into account only terms with nonnegative indices.

{\bf Theorem B:$\;$} 
{\sl Let the function $\:c(\varepsilon)\:$ satisfy the condition
\begin{eqnarray}
\lim_{\varepsilon \rightarrow 0} \varepsilon c^3(\varepsilon) = 0
\nonumber
\end{eqnarray}
Then for arbitrary initial points $\;x_0$, $\;z_0$, $\;\vec{y}_0\;$,
and for arbitrary nonnegative integer $\;N\;$ and  positive $\;L$
\begin{eqnarray}
\lim_{\varepsilon \rightarrow 0} 
\hspace{0.3cm}
\max_{0 \leq t \leq L / \varepsilon^2}
\hspace{0.3cm}
\left|
\left<
\bar{\cal M}_N^{-1} (\varepsilon^2 s_t^{\varepsilon})
\:\vec{\cal V}(I_{m,\:k}(s_t^{\varepsilon});\,N)
 \;-\;  
\vec{\cal V}(I_{m,\:k}(0);\,N)  \right> \right| = 0
\nonumber 
\end{eqnarray} 
where 
the matrix $\;\bar{\cal M}_N(\tau)\;$ is the fundamental
matrix 
solution of the system of linear
ordinary differential equations with constant coefficients
(\ref{srk}).
}

{\bf Remark 1:$\;$} For further purposes it is important
to note that the statement of the theorem B can also be written 
in the form 

\begin{eqnarray}
\lim_{\varepsilon \rightarrow 0} 
\hspace{0.3cm}
\max_{0 \leq t \leq L / \varepsilon^2}
\hspace{0.3cm}
\left|
\left<
\vec{a}_N(\varepsilon^2 s_t^{\varepsilon})\:
\cdot 
\vec{\cal V}(I_{m,\:k}(s_t^{\varepsilon}),\,N)
 \;-\;
\vec{a}_N(0) \cdot 
\vec{\cal V}(I_{m,\:k}(0),\,N) \right> \right| = 0
\nonumber 
\end{eqnarray} 
where $\;\vec{a}_N(\tau)\;$ is an arbitrary 
$\;(N+1)(N+2)/2$-dimensional vector
satisfying
\begin{eqnarray}
\frac{d \vec{a}_N}{d \tau}\:\;=\;-
\bar{\cal K}_{N}^{\top} \;\vec{a}_N 
\label{srk99}
\end{eqnarray}

{\bf Remark 2:$\;$} For physical applications one can
neglect the small difference
between $t$ and $s_t^{\varepsilon}$ 
(see theorem A)and we have

\begin{eqnarray}
\left<
\vec{\cal V}(I_{m,\:k}(t),\,N)
\right>\,\,
\approx 
\,\,
\bar{\cal M}_N (\varepsilon^2 t)\,\,
\vec{\cal V}(I_{m,\:k}(0),\,N) 
\nonumber 
\end{eqnarray} 

\section{Nonresonant Case}

\hspace*{0.5cm}
Let us now define what we mean by nonresonant.

{\bf Definition:$\;$}
{\sl 
We shall say that there are no resonances of
order $\:m \:\geq\: 0\:$ if for all integers $p$, $q$ 
such that $\;(p,\:q)\:\in\: F \times F\;$
\begin{eqnarray}
m \:\omega_0 \;\neq\; \nu_p \:+\: \nu_q
\nonumber
\end{eqnarray}
}

{\bf Definition:$\;$}
{\sl 
We shall say that there are no resonances up to
order $\:m \:\geq\: 0\:$ if for all integers $p$, $q$ 
such that $\;(p,\:q)\:\in\: F \times F\;$
\begin{eqnarray}
k \:\omega_0 \;\neq\; \nu_p \:+\: \nu_q
\hspace{0.4cm}
\mbox{for}
\hspace{0.4cm}
k \:=\: 1,\ldots,m
\nonumber
\end{eqnarray}
}

In the nonresonant case only the values of
$\;\bar{c}_6(m, \:k)\;$ and
$\;\bar{c}_7(m, \:k)\;$ will be different 
from zero.
Introduce for them special notations  
\begin{eqnarray}
{\cal A}_{m, \:k} \;=\;
\bar{c}_7(m, \:k),
\hspace*{1.0cm}
{\cal C}_{m, \:k}\;=\; 
\bar{c}_6(m, \:k) 
\nonumber
\end{eqnarray}
Note that $\;{\cal C}_{m,\:k}\;$ is a symmetrical 
function of its arguments, i.e.
$\;{\cal C}_{m,\:k}\:=\:{\cal C}_{k,\:m}\;$
and it is also a real valued function 
 i.e.
$\;{\cal C}_{m,\:k}\:=\:{\cal C}_{m,\:k}^*$, and
the function $\;{\cal A}_{m,\:k}\;$ satisfies
$\;{\cal A}_{m,\:k}\:=\:{\cal A}_{k,\:m}^*$.

For the following let us also introduce special
notations  for the real and imaginary parts of
$\;{\cal A}_{m,\:k}\;$ 
\begin{eqnarray}
\bar{\cal A}_{m,\:k} \;=\; 
\frac{{\cal A}_{m,\:k} \:+\:{\cal A}_{k,\:m}}{2},
\hspace{1.0cm}
\bar{\cal B}_{m,\:k} \;=\; 
\frac{{\cal A}_{m,\:k} \:-\:{\cal A}_{k,\:m}}{2i}
\nonumber
\end{eqnarray}
We shall call $\bar{\cal A}_{m,\:k}$ 
and $\bar{\cal B}_{m,\:k}$ for reasons which will 
become clear later 
diffusion coefficient and tune shift respectively.
Note that 
$\bar{\cal A}_{m,\:m} \:=\: {\cal A}_{m,\:m}\;$ and 
$\bar{\cal B}_{m,\:m} \:=\:0$.

{\bf Theorem C:$\;$} 
{\sl
Let there be no resonances up to order $4$
and let the function $\:c(\varepsilon)\:$ satisfy the condition
\begin{eqnarray}
\lim_{\varepsilon \rightarrow 0} \varepsilon c^3(\varepsilon) = 0
\nonumber
\end{eqnarray}
Then for any initial points $\;x_0$, $\;z_0$, $\;\vec{y}_0\;$,
for any nonnegative integers $\;m\;$, $\;k\;$ 
and for any positive $\;L$
\begin{eqnarray}
\lim_{\varepsilon \rightarrow 0} 
\hspace{0.3cm}
\max_{0 \leq t \leq L / \varepsilon^2}
\hspace{0.3cm}
\left|
\left<
\sum \limits_{p = 0}^{q}
a_p^{m,\:k}(\varepsilon^2 s_t^{\varepsilon})
I_{m-p,\:k-p}(s_t^{\varepsilon}) \:-\:
\sum \limits_{p = 0}^{q}
a_p^{m,\:k}(0)
I_{m-p,\:k-p}(0) 
\right>
\right| \;=\; 0
\nonumber
\end{eqnarray}
where $\;q \:=\: \min\{m,\, k\}\;$ and the functions
$\;a_p^{m,\:k}(\tau)\;$ are an arbitrary solution of the system of
linear ordinary differential equations with constant coefficients
\begin{eqnarray}
\frac{d a_0^{m,\:k}}{d \tau}\;=\;- {\cal A}_{m,\:k} \: a_0^{m,\:k}
\nonumber
\end{eqnarray}
\begin{eqnarray}
\frac{d a_p^{m,\:k}}{d \tau}\;=\;- {\cal A}_{m-p,\:k-p} \: a_p^{m,\:k}
\;-\; {\cal C}_{m-p+1,\:k-p+1}\: a_{p-1}^{m,\:k}
\nonumber
\end{eqnarray}
\begin{eqnarray}
p \;=\; 1, \ldots, q
\nonumber
\end{eqnarray}
}

The proof of this theorem can be obtained from the remark to
the theorem B with help of some 
straightforward calculations.

{\bf Remark 1:$\;$} We would like to note that for 
the study of the behaviour of first order moments 
(i.e. when $\;m\:+\:k\:=\:1$) we actually need to avoid
resonances in theorem B up to order 2 only.
  
{\bf Remark 2:$\;$} The general solution of the system of
differential equations for the coefficients $a_{p}^{m,\:k}$
has the form
\begin{eqnarray}
a_{0}^{m,\:k}(\tau)
\;=\;
a_{0}^{m,\:k}(0) 
\cdot \exp\left(-{\cal A}_{m,\:k} \,\tau\right)
\nonumber
\end{eqnarray}
\begin{eqnarray}
a_{p}^{m,\:k}(\tau)
\;=\;
\left(
a_{p}^{m,\:k}(0)-
{\cal C}_{m-p+1,\:k-p+1}
\int \limits_0^{\tau}
a_{p-1}^{m,\:k}(\zeta) 
\cdot \exp\left({\cal A}_{m-p,\:k-p} \,\zeta\right)
d \zeta
\right)\cdot 
\nonumber
\end{eqnarray}
\begin{eqnarray}
\cdot \exp\left(-{\cal A}_{m-p,\:k-p} \,\tau\right)
\nonumber
\end{eqnarray}
\begin{eqnarray}
p \;=\; 1,\,\ldots,\,q
\nonumber
\end{eqnarray}
Choosing the initial conditions 
\begin{eqnarray}
a_{0}^{m,\:k}(0) \;=\;1,
\hspace*{1.0cm}
a_{p}^{m,\:k}(0) \;=\;0,
\hspace*{1.0cm}
p \;=\; 1,\,\ldots,\,q
\nonumber
\end{eqnarray}
the statement of the theorem C can be rewritten in the form
\begin{eqnarray}
\lim_{\varepsilon \rightarrow 0} 
\hspace{0.3cm}
\max_{0 \leq t \leq L / \varepsilon^2}
\hspace{0.3cm}
\nonumber
\end{eqnarray}
\begin{eqnarray}
\left|
\left<
\exp(-\varepsilon^2 {\cal A}_{m,\:k} s_t^{\varepsilon})
I_{m,\:k}(s_t^{\varepsilon}) 
-
\left(
I_{m,\:k}(0) -
\sum \limits_{p = 1}^{q}
a_p^{m,\:k}(\varepsilon^2 s_t^{\varepsilon})
I_{m-p,\:k-p}(s_t^{\varepsilon})
\right) 
\right>
\right| = 0
\nonumber
\end{eqnarray}

In the case when $\;m\,\neq\,k\;$ we can use the
real valued functions $\:U_{m,\:k}\:$ and
$\:V_{m,\:k}\:$ instead of the complex valued $\:I_{m,\:k}\,$.
Due to the symmetries
$\;U_{m,\:k}\:=\:U_{k,\:m}\;$ and
$\;V_{m,\:k}\:=\:-\,V_{k,\:m}\;$ 
it is enough to consider only the case when
$\;m\:>\:k\,$. So we have
  
{\bf Corollary 1:$\;$} 
{\sl
Let there be no resonances up to order $4$
and let the function $\:c(\varepsilon)\:$ satisfy the condition
\begin{eqnarray}
\lim_{\varepsilon \rightarrow 0} \varepsilon c^3(\varepsilon) = 0
\nonumber
\end{eqnarray}
Then for any initial points $\;x_0$, $\;z_0$, $\;\vec{y}_0\;$,
for any nonnegative integers $\;m,\;k\;$ satisfying $\;m\:>\:k\;$,
and for any positive $\;L$
\begin{eqnarray}
\lim_{\varepsilon \rightarrow 0} 
\hspace{0.3cm}
\max_{0 \leq t \leq L / \varepsilon^2}
\hspace{0.3cm}
\left|
\left<
\sum \limits_{p = 0}^{k}
M_p^{m,\:k}(\varepsilon^2 s_t^{\varepsilon})
\cdot
\vec{W}_p^{m,\:k}(s_t^{\varepsilon}) 
-
\sum \limits_{p = 0}^{k}
M_p^{m,\:k}(0)
\cdot
\vec{W}_p^{m,\:k}(0) 
\right>
\right| 
= 0
\nonumber
\end{eqnarray}
where
\begin{eqnarray}
\vec{W}_p^{m,\:k}(\tau) =
\left(
\begin{array}{c}
U_{m-p,\:k-p}(\tau) \\
\\
V_{m-p,\:k-p}(\tau) 
\end{array}
\right),
\hspace{0.8cm}
M_p^{m,\:k}(\tau) =
\left(
\begin{array}{rr}
\alpha_p^{m,\:k}(\tau) & -\beta_p^{m,\:k}(\tau) \\
\\
\beta_p^{m,\:k}(\tau) &  \alpha_p^{m,\:k}(\tau) 
\end{array}
\right)
\nonumber
\end{eqnarray}
and the functions
$\;\alpha_p^{m,\:k}(\tau)\;$ and 
$\;\beta_p^{m,\:k}(\tau)\;$ 
are an arbitrary real solution of the system of
linear ordinary differential equations
with constant coefficients
\begin{eqnarray}
\frac{d}{d \tau}
\left(
\begin{array}{c}
\alpha_0^{m,\:k} \\
\\
\beta_0^{m,\:k} 
\end{array}
\right)
=
\bar{R}_0^{m,\:k}
\left(
\begin{array}{c}
\alpha_0^{m,\:k} \\
\\
\beta_0^{m,\:k} 
\end{array}
\right)
\nonumber
\end{eqnarray}

\begin{eqnarray}
\frac{d}{d \tau}
\left(
\begin{array}{c}
\alpha_p^{m,\:k} \\
\\
\beta_p^{m,\:k} 
\end{array}
\right)
=
\bar{R}_p^{m,\:k}
\left(
\begin{array}{c}
\alpha_p^{m,\:k} \\
\\
\beta_p^{m,\:k} 
\end{array}
\right)
-{\cal C}_{m-p+1,\:k-p+1}
\left(
\begin{array}{c}
\alpha_{p-1}^{m,\:k} \\
\\
\beta_{p-1}^{m,\:k} 
\end{array}
\right)
\nonumber
\end{eqnarray}
\begin{eqnarray}
p \;=\; 1, \ldots, k
\nonumber
\end{eqnarray}
\begin{eqnarray}
\bar{R}_p^{m,\:k}
=
\left(
\begin{array}{rr}
-\bar{\cal A}^{m-p,\:k-p} & \bar{\cal B}^{m-p,\:k-p} \\
\\
-\bar{\cal B}^{m-p,\:k-p} & -\bar{\cal A}^{m-p,\:k-p} \\
\end{array}
\right)
\nonumber
\end{eqnarray}
}

For the important particular case when we do not have an external
noise in our system, i.e. $\xi(t)\:\equiv\:0$ 
(that means that we can put $\vec{h}(t)\:\equiv\:\vec{0}$ and
hence 
all ${\cal C}_{m,\:k}\:=\:0$)
the differential equations defining the functions
$\;a_p^{m,\:m}$, $\;\alpha_p^{m,\:k}\;$ and $\;\beta_p^{m,\:k}\;$ 
admit the simple solution
\begin{eqnarray} 
a_0^{m,\:m}(\tau)\;=\;
\exp(-\bar{\cal A}_{m,\:m} \tau)
\:a_0^{m,\:m}(0)
\nonumber
\end{eqnarray}
\begin{eqnarray}
\left(
\begin{array}{c}
\alpha_0^{m,\:k}(\tau) \\
\\
\beta_0^{m,\:k}(\tau) 
\end{array}
\right)
=
\exp(-\bar{\cal A}_{m,\:k} \tau)
\left(
\begin{array}{rr}
\cos(\bar{\cal B}_{m,\:k}\tau) &
\sin(\bar{\cal B}_{m,\:k}\tau)\\
\\
-\sin(\bar{\cal B}_{m,\:k}\tau) &
\cos(\bar{\cal B}_{m,\:k}\tau)
\end{array}
\right)
\left(
\begin{array}{c}
\alpha_0^{m,\:k}(0) \\
\\
\beta_0^{m,\:k}(0) 
\end{array}
\right)
\nonumber
\end{eqnarray}
\begin{eqnarray}
a_p^{m,\:m}(\tau)\:\equiv\:0, 
\hspace{0.7cm} 
\alpha_p^{m,\:k}(\tau)\:\equiv\:0, 
\hspace{0.7cm}
\beta_p^{m,\:k}(\tau)\:\equiv\:0, 
\hspace{0.7cm}
p \neq 0 
\nonumber
\end{eqnarray}
Choosing initial conditions
\begin{eqnarray}
a_0^{m,\:m}(0)\:=\:1,
\hspace{1.0cm} 
\alpha_0^{m,\:k}(0)\:=\:1, 
\hspace{1.0cm}
\beta_0^{m,\:k}(0)\:=\:0
\nonumber
\end{eqnarray}
we get the following

{\bf Corollary 2:$\;$} 
{\sl
Let $\;\xi(t)\:\equiv\:0\;$ and let
there be no resonances of orders $2$ and $4$,
and let the function $\:c(\varepsilon)\:$ satisfy the condition
\begin{eqnarray}
\lim_{\varepsilon \rightarrow 0} \varepsilon c^3(\varepsilon) = 0
\nonumber
\end{eqnarray}
Then for any initial points $\;x_0$, $\;z_0$, $\;\vec{y}_0\;$,
for any nonnegative integers $\;m,\;k\;$ satisfying 
$\;m\:\geq\:k\;$,
and for any positive $\;L$
\begin{center}
\begin{eqnarray}
\lim_{\varepsilon \rightarrow 0} 
\hspace{0.2cm}
\max_{0 \leq t \leq L / \varepsilon^2}
\hspace{0.3cm}
\left|
\left<
\exp\left(-\varepsilon^2 \:\bar{\cal A}_{m,\:m}\:s_t^{\varepsilon}\right)
r^m \left(s_t^{\varepsilon}\right) 
- r^m(0)
\right>
\right| = 0
\nonumber
\end{eqnarray}
\vspace*{0.2cm}
for $\;k \:=\:m\;$ and
\vspace*{0.2cm}
\begin{eqnarray}
\lim_{\varepsilon \rightarrow 0} 
%\hspace{0.3cm}
\max_{0 \leq t \leq L / \varepsilon^2}
%\hspace{0.3cm}
\left|
\left<
\exp\left(-\varepsilon^2 \:\bar{\cal A}_{m,k}\: s_t^{\varepsilon}\right)
%\cdot 
\bar{M}_m^k(s_t^{\varepsilon})
%\cdot
\left(
\begin{array}{c}
\bar{U}_{m,k}(s_t^{\varepsilon}) \\
\\
\bar{V}_{m,k}(s_t^{\varepsilon}) 
\end{array}
\right)
-
\left(
\begin{array}{c}
\bar{U}_{m,k}(0) \\
\\
\bar{V}_{m,k}(0) 
\end{array}
\right)
\right>
\right| = 0
\nonumber
\end{eqnarray}
where
\begin{eqnarray}
\bar{M}_m^k(\tau) =
\left(
\begin{array}{rr}
 \cos\left(\Delta_{m,k}^{\varepsilon}\tau\right) &
-\sin\left(\Delta_{m,k}^{\varepsilon}\tau\right) \\
\\
 \sin\left(\Delta_{m,k}^{\varepsilon}\tau\right) &
 \cos\left(\Delta_{m,k}^{\varepsilon}\tau\right)
\end{array}
\right),
\hspace{0.25cm}
\Delta_{m,k}^{\varepsilon} =
(m-k)\:\omega_0 - \varepsilon^2 \bar{\cal B}_{m,k}
\nonumber
\end{eqnarray}
\vspace*{0.4cm}
otherwise.
\end{center}
}

For the important case of constant vectors 
$\;\vec{b}\;$, $\;\vec{h}\;$ and $\;\vec{d}\;$
the formulae for $\;\bar{{\cal A}}_{m,\:k}$,
$\;\bar{{\cal B}}_{m,\:k}\;$ and $\;{\cal C}_{m,\:k}\;$
take the simplified form
\begin{eqnarray}
\bar{{\cal A}}_{m,\:k} \;=\; -\frac{m+k}{2}\:\alpha \:-\: 
\frac{(m-k)^2}{4} \Psi_{c}(0)\: \vec{b}\cdot \vec{b} 
\:+\:\frac{(m+k)^2}{4} \Psi_{c}(0)\: \vec{d}\cdot \vec{d} \:+ 
\nonumber
\end{eqnarray}
\begin{eqnarray}
+\frac{m + 2 m k + k}{4} 
\left[
\rule[0.25cm]{0cm}{0.25cm}
\Psi_{c}(2 \omega_0)\: \vec{b} \cdot \vec{b} +
\Psi_{c}(2 \omega_0)\: \vec{d} \cdot \vec{d} +
\left(
\Psi_{s}(2 \omega_0)-\Psi_{s}^{\top}(2 \omega_0)
\right)\:
\vec{d} \cdot \vec{b}\: 
\right]
\nonumber
\end{eqnarray}

\begin{eqnarray}
\bar{{\cal B}}_{m,\:k} \;=\;
\frac{m^2-k^2}{4}
\left(\Psi_{c}(0)\:+\:\Psi_{c}^{\top}(0)\right)
\vec{d} \cdot \vec{b} \:+ 
\nonumber
\end{eqnarray}
\begin{eqnarray}
+\:\frac{m-k}{4}
\left[
\rule[0.25cm]{0cm}{0.25cm}
\Psi_{s}(2 \omega_0)\:\vec{b} \cdot \vec{b}\:+\:
\Psi_{s}(2 \omega_0)\:\vec{d} \cdot \vec{d}
\:-\:
\left(\Psi_{c}(2 \omega_0)\:-\:\Psi_{c}^{\top}(2 \omega_0)\right)
\vec{d} \cdot \vec{b} 
\:\right]
\nonumber
\end{eqnarray}

\begin{eqnarray}
{\cal C}_{m,\:k} \;=\;
\frac{m k}{2}\:\Psi_{c}(\omega_0)\:\vec{h} \cdot \vec{h}
\nonumber
\end{eqnarray}

\section{First and Second Order Moments}

First order moments in the nonresonant case:

{\bf Corollary C1:$\;$}
{\sl
Let there be no resonances up to order $2$ and 
let the function $\:c(\varepsilon)\:$ satisfy the condition
\begin{eqnarray}
\lim_{\varepsilon \rightarrow 0} \varepsilon c^3(\varepsilon) = 0
\nonumber
\end{eqnarray}
Then for any initial points $\;x_0$, $\;z_0$, $\;\vec{y}_0\;$ 
and for any positive $\;L$
\begin{eqnarray}
\lim_{\varepsilon \rightarrow 0} 
\hspace{0.3cm}
\max_{0 \leq t \leq L / \varepsilon^2}
\hspace{0.3cm}
\left|
\left<
\exp\left(-\varepsilon^2 \bar{\cal A}_{1,\:0} \:s_t^{\varepsilon}\right)
\: M(s_t^{\varepsilon})\:
\left(
\begin{array}{c}
x(s_t^{\varepsilon}) \\
\\
z(s_t^{\varepsilon}) 
\end{array}
\right)
-
\left(
\begin{array}{c}
x_0 \\
\\
z_0 
\end{array}
\right)
\right>
\right| = 0
\nonumber
\end{eqnarray}
where
\begin{eqnarray}
M(\tau) =
\left(
\begin{array}{rr}
 \cos\left(\left(\omega_0 \:-\: \varepsilon^2 \bar{\cal B}_{1,\:0}\right)
\tau\right) &
-\sin\left(\left(\omega_0 \:-\: \varepsilon^2 \bar{\cal B}_{1,\:0}\right)
\tau\right) \\
\\
 \sin\left(\left(\omega_0 \:-\: \varepsilon^2 \bar{\cal B}_{1,\:0}\right)
\tau\right) &
 \cos\left(\left(\omega_0 \:-\: \varepsilon^2 \bar{\cal B}_{1,\:0}\right)
\tau\right)
\end{array}
\right)
\nonumber
\end{eqnarray}
}

Second order moments in nonresonant case:

{\bf Corollary C2:$\;$}
{\sl
Let there be no resonances up to order $4$ and 
let the function $\:c(\varepsilon)\:$ satisfy the condition
\begin{eqnarray}
\lim_{\varepsilon \rightarrow 0} \varepsilon c^3(\varepsilon) = 0.
\nonumber
\end{eqnarray}
Then for any initial points 
$x_0$, $z_0$, $\vec{y}_0 \in R^n$ and for any positive $L$
\begin{center}
\begin{eqnarray}
\lim_{\varepsilon \rightarrow 0} 
\hspace{0.2cm}
\max_{0 \leq t \leq L / \varepsilon^2}
\hspace{0.3cm}
\left|
\left<
r \left(s_t^{\varepsilon}\right) 
- r_0 - \varepsilon^2 \:2\: {\cal C}_{1,\:1} s_t^{\varepsilon}
\right>
\right| = 0
\nonumber
\end{eqnarray}
\vspace*{0.2cm}
for $\;\bar{{\cal A}}_{1,\:1} \;=\; 0\;$ and
\vspace*{0.2cm}
\begin{eqnarray}
\lim_{\varepsilon \rightarrow 0} 
\hspace{0.3cm}
\max_{0 \leq t \leq L / \varepsilon^2}
\hspace{0.3cm}
\left|
\left<
\left(r \left(s_t^{\varepsilon}\right) + 
\frac{2\:{\cal C}_{1,\:1}}{\bar{\cal A}_{1,\:1}}\right)
\exp\left(-\varepsilon^2 \bar{\cal A}_{1,\:1} s_t^{\varepsilon}\right)
-\left(r_0 + \frac{2\:{\cal C}_{1,\:1}}{\bar{\cal A}_{1,\:1}} \right)
\right>
\right| = 0
\nonumber
\end{eqnarray}
\vspace*{0.4cm}
otherwise.
\end{center}
}

To estimate the behaviour of the remainder of the second moments 
we shall use the functions
\begin{eqnarray}
\bar{U}_{2,\:0} \; = \; \frac{x^2 - z^2}{4}
\hspace{1.0cm}
\bar{V}_{2,\:0} \; = \; \frac{x z}{2}
\nonumber
\end{eqnarray}

{\bf Corollary C3:}
{\sl
Let there be no resonances up to order $4$ and 
let the function $\:c(\varepsilon)\:$ satisfy the condition
\begin{eqnarray}
\lim_{\varepsilon \rightarrow 0} \varepsilon c^3(\varepsilon) = 0
\nonumber
\end{eqnarray}
Then for any initial points $\;x_0$, $\;z_0$, $\;\vec{y}_0\;$ 
and for any positive $\;L$
\begin{eqnarray}
\lim_{\varepsilon \rightarrow 0} 
\hspace{0.3cm}
\max_{0 \leq t \leq L / \varepsilon^2}
%\hspace{0.3cm}
\left|
\left<
\exp\left(-\varepsilon^2 \:\bar{\cal A}_{2,\:0}\: s_t^{\varepsilon}\right)
\: M(s_t^{\varepsilon})\:
\left(
\begin{array}{c}
\bar{U}_{2,\:0}(s_t^{\varepsilon}) \\
\\
\bar{V}_{2,\:0}(s_t^{\varepsilon}) 
\end{array}
\right)
-
\left(
\begin{array}{c}
\bar{U}_{2,\:0}(0) \\
\\
\bar{V}_{2,\:0}(0) 
\end{array}
\right)
\right>
\right| = 0
\nonumber
\end{eqnarray}
where
\begin{eqnarray}
M(\tau) =
\left(
\begin{array}{rr}
 \cos\left(\left(2\omega_0 - \varepsilon^2 \bar{\cal B}_{2,\:0}\right)
\tau\right) &
-\sin\left(\left(2\omega_0 - \varepsilon^2 \bar{\cal B}_{2,\:0}\right)
\tau\right) \\
\\
 \sin\left(\left(2\omega_0 - \varepsilon^2 \bar{\cal B}_{2,\:0}\right)
\tau\right) &
 \cos\left(\left(2\omega_0 - \varepsilon^2 \bar{\cal B}_{2,\:0}\right)
\tau\right)
\end{array}
\right)
\nonumber
\end{eqnarray}
}

\section{Comparison with White Noise Model}

\hspace*{0.5cm}
As a special case we consider white noise in
this chapter i.e.
\begin{eqnarray}
\vec{y} = C \, \dot{\vec{w}}(t)
\label{wn9}
\end{eqnarray}
where $\;C\;$ is a real constant ($n \times r$) matrix 
and $\;\vec{w}(t)\;$ 
is an $r$-dimensional Brownian motion. Substituting
(\ref{wn9}) into (\ref{m02})
we have 
\begin{eqnarray}
\left\{
\begin{array}{l}
d x \;=\; \hspace{0.3cm} \omega_0 \: z \: d t\\
\\
d z \;=\; -\omega_0 \: x \: d t \:-\:
\varepsilon^2 \:\alpha\: z \:d t \:+\:
\varepsilon \: C^{\top}
\left(\vec{h} - z \vec{d} - x \vec{b} \right) 
\cdot d \vec{w}(t)
\end{array}
\right.
\label{m02wn}
\end{eqnarray}
As usual for the case of multiplicative noise we shall
treat the system (\ref{m02wn}) as a system of 
Stratonovich's stochastic differential equations.

Introduce the matrix $\:\Phi\:=\:\frac{1}{2} C C^{\top}\:$ 
which plays the role of the spectral density for the noise
model (\ref{wn9})
and define functions 
$\:\breve{c}_l(m,\:k)\:$ with the help of 

\begin{eqnarray} 
\breve{c}_1(m, \:k) \;=\;
\frac{m}{2}
\sum \limits_{\nu_l-\nu_p\:=\:\omega_0}
\left\{
\rule[0.2cm]{0cm}{0.2cm}
(m-2k-1)\: \Phi \:\vec{h}_p \cdot \vec{b}_l 
\;-\;
i\:(m+2k)\: \Phi \:\vec{h}_p \cdot \vec{d}_l 
\right\},
\nonumber
\end{eqnarray}

\begin{eqnarray} 
\breve{c}_2(m, \:k) \;=\;
-\frac{m(m-1)}{4}
\sum \limits_{\nu_l-\nu_p\:=\:2\omega_0}
\Phi \:\vec{h}_p \cdot \vec{h}_l, 
\nonumber
\end{eqnarray}

\begin{eqnarray} 
\breve{c}_3(m, \:k) \;=\;
\frac{m(m-1)}{2}
\sum \limits_{\nu_l-\nu_p\:=\:3\omega_0}
\Phi \:\vec{h}_p \cdot 
\left(\vec{b}_l - i \vec{d}_l\right), 
\nonumber
\end{eqnarray}

\begin{eqnarray} 
\breve{c}_4(m, \:k) \;=\;
-\frac{m(m-1)}{4}
\sum \limits_{\nu_l-\nu_p\:=\:4\omega_0}
\Phi \:
\left(\vec{b}_p + i \vec{d}_p\right) 
\cdot 
\left(\vec{b}_l - i \vec{d}_l\right), 
\nonumber
\end{eqnarray}

\begin{eqnarray} 
\breve{c}_5(m, \:k) \;=\;
\frac{m}{2}
\sum \limits_{\nu_l-\nu_p\:=\:2\omega_0}
\left\{
\rule[0.2cm]{0cm}{0.2cm}
-(m+k)\: \Phi \:\vec{d}_p \cdot \vec{d}_l 
\;+\;
\right.
\nonumber
\end{eqnarray}
\begin{eqnarray} 
\left.
+\;(k-m+1)\: \Phi \:\vec{b}_p \cdot \vec{b}_l 
\;+\;
i\:(2k+1)\: \Phi \:\vec{b}_p \cdot \vec{d}_l 
\rule[0.2cm]{0cm}{0.2cm}
\right\},
\nonumber
\end{eqnarray}

\begin{eqnarray} 
\breve{c}_6(m, \:k) \;=\;
\frac{m k}{2}
\sum \limits_{p\:=\:-\infty}^{\infty}
\Phi \:\vec{h}_p \cdot \vec{h}_p, 
\nonumber
\end{eqnarray}

\begin{eqnarray} 
\breve{c}_7(m, \:k) \;=\;
-\frac{m+k}{2}\:\alpha \;+\;
\frac{4mk-m(m-1)-k(k-1)}{4}
\sum \limits_{p\:=\:-\infty}^{\infty}
\Phi \:\vec{b}_p \cdot \vec{b}_p 
\;+\;
\nonumber
\end{eqnarray}

\begin{eqnarray} 
+\:\frac{4mk+m(m+1)+k(k+1)}{4}
\sum \limits_{p\:=\:-\infty}^{\infty}
\Phi \:\vec{d}_p \cdot \vec{d}_p 
\;+\;
i\:\frac{m^2-k^2}{2}
\sum \limits_{p\:=\:-\infty}^{\infty}
\Phi \:\vec{d}_p \cdot \vec{b}_p. 
\nonumber
\end{eqnarray}

Consider now the system of ordinary differential
equations with constant coefficients
\begin{eqnarray}
\frac{d}{d \tau}\:\vec{\cal V}(a_{m,\:k};\,N)\;=\;
\breve{\cal K}_{N} \; \vec{\cal V}(a_{m,\:k};\,N)
\label{srk_breve}
\end{eqnarray}
generated with the help of the rule
\begin{eqnarray}
\begin{array}{llllll}
\frac{d}{d \tau}\: a_{m,\:k}& =
&
\breve{c}_2(m,\:k) \: a_{m-2,\:k} & + &
\breve{c}_2^{\;*}(k,\:m) \: a_{m,\:k-2} & + \\
\\
& & 
\breve{c}_1(m,\:k) \: a_{m-1,\:k} & + &
\breve{c}_1^{\;*}(k,\:m) \: a_{m,\:k-1} & + \\
\\
& & 
\breve{c}_3(m,\:k) \: a_{m-2,\:k+1} & + &
\breve{c}_3^{\;*}(k,\:m) \: a_{m+1,\:k-2} & + \\
\\
& &
\breve{c}_4(m,\:k) \: a_{m-2,\:k+2} & + &
\breve{c}_4^{\;*}(k,\:m) \: a_{m+2,\:k-2} & + \\
\\
& & 
\breve{c}_5(m,\:k) \: a_{m-1,\:k+1}  & + &
\breve{c}_5^{\;*}(k,\:m) \: a_{m+1,\:k-1} & + \\
\\
& & 
\breve{c}_6(m,\:k) \: a_{m-1,\:k-1}  & + &
\breve{c}_7(m,\:k) \: a_{m,\:k}
\end{array}
\label{rule_r_breve}
\end{eqnarray}
where on the right hand side of (\ref{rule_r_breve}) we take
into account only terms with nonnegative indices.

{\bf Theorem D:$\;$} 
{\sl 
For any initial points $\;x_0$, $\;z_0$, $\;\vec{y}_0\;$,
for any nonnegative integer $\;N\;$ and for any positive $\;L$
\begin{eqnarray}
\lim_{\varepsilon \rightarrow 0} 
\hspace{0.3cm}
\max_{0 \leq t \leq L / \varepsilon^2}
\hspace{0.3cm}
\left|
\left<
\breve{\cal M}_N^{-1} (\varepsilon^2 t)
\:\vec{\cal V}(I_{m,\:k}(t);\,N)
 \;-\;  
\vec{\cal V}(I_{m,\:k}(0);\,N)  \right> \right| = 0
\nonumber 
\end{eqnarray} 
where 
the matrix $\;\breve{\cal M}_N(\tau)\;$ is the fundamental
matrix 
solution of the system of linear
ordinary differential equations with constant coefficients
(\ref{srk_breve}).
}

Note that in this case and for the noise model
introduced below we have not to distinguish
between  $s_t^{\varepsilon}$ and $t$.
We also mention that
if we substitute into the expressions 
of $\:\bar{c}_l(m,\:k)\:$ 
the matrix $\:\Phi\:$ instead of the matrix $\:\Psi\:$
("spectral density" of white noise)
we exactly get  $\:\breve{c}_l(m,\:k)\:$.

\section{Another  Noise Model}

The technique derived in this paper
can be applied to
a wide class of noise models.
As a model of noise in this section 
we consider the stochastic
processes represented by the following
trigonometrical polynomials\footnote{In order 
 not to deal with conditions
similar to (\ref{cond1}) and (\ref{cond2}) we 
consider the case of a finite trigonometrical sum.
The extension to the case of infinite series and
also the proof of the theorem E we leave 
as an exercise for the interested reader.}
(cosine and sine functions with random phases)
\begin{eqnarray}
\eta(t) \;=\; 
\sum \limits_{m = -q}^{q}
\eta_{m} \exp\left(i\left(\nu_m t \:+\:
\vec{v}_m \cdot \vec{y}\,\right)\right),
\hspace{1.0cm} 
\eta_{-m} = \left(\eta_{m}\right)^{*}
\nonumber
\end{eqnarray}
\begin{eqnarray}
\xi(t) \;=\; 
\sum \limits_{m = -q}^{q}
\xi_{m} \exp\left(i\left(\nu_m t \:+\:
\vec{v}_m \cdot \vec{y}\,\right)\right),
\hspace{1.0cm} 
\xi_{-m} = \left(\xi_{m}\right)^{*}
\nonumber
\end{eqnarray}
\begin{eqnarray}
\gamma(t) \;=\; 
\sum \limits_{m = -q}^{q}
\gamma_{m} \exp\left(i\left(\nu_m t \:+\:
\vec{v}_m \cdot \vec{y}\,\right)\right),
\hspace{1.0cm} 
\gamma_{-m} = \left(\gamma_{m}\right)^{*}
\nonumber
\end{eqnarray}
with real $\;\nu_m\;$ and $\;\vec{v}_m\:\in\:R^n\;$ 
satisfying 
the conditions 
\begin{eqnarray}
|\nu_{l}\:+\:\nu_{m}|\:+\:
|\vec{v}_m \:+\:\vec{v}_l| \:=\:0 \;\Leftrightarrow\; 
m\:+\:l\:=\:0
\nonumber
\end{eqnarray}
where the integers $\;m,\,l\:$ obey $\;m,\,l\:=\:-q,\,\ldots,\,q$.

The vector $\;\vec{y}\:\in\:R^n\;$ is assumed to be a
solution of the following Ito's system
\begin{eqnarray}
d\,\vec{y}\;=\;\sqrt{2}\,B\,d \vec{w}(t) 
\nonumber
\end{eqnarray}
where $\:B\:$ is a real constant $(n \times r)$ matrix
and $\:\vec{w}(t)\:$ is an $r$-dimensional Brownian motion.
For simplicity we assume that the $(n \times n)$ matrix
$\:B\,B^{\top}\:$ is nondegenerate and 
$\:|\,\vec{v}_m\,|\,\neq\,0\:$ for all $\:m\,=\,-q,\ldots,q\:$
(i.e. we do not have deterministic harmonics in our
perturbation model). 

For $\;p \:=\:-q,\,\ldots,\,q\;$ we introduce real vectors 
$\;\vec{u}_p\:=\:B^{\top}\vec{v}_p\:\in\:R^r\:$ which
satisfy $\;|\vec{u}_p|\:\neq\:0$, and a function 
$\:\Omega\left(\omega,\, \vec{u}_p\right)$
\begin{eqnarray}
\Omega\left(\omega,\, \vec{u}_p\right)\;=\;
\frac{\;|\vec{u}_p|^2\:+\:i\,\omega\;}
{\;|\vec{u}_p|^4 \:\;+\:\;\omega^2\;}
\nonumber
\end{eqnarray}
and define $\:\tilde{c}_l(m,\,k)\:$ as follow

\begin{eqnarray}
\tilde{c}_1(m,\,k)\;=\; \frac{m}{4}
\left(
\sum
\limits_{\stackrel{|\nu_p+\nu_l+\omega_0|+}{+|\vec{v}_p+\vec{v}_l|=0}}
\left\{
(m-1)\:
\Omega\left(\omega_0\,+\,\nu_p,\, \vec{u}_p\right)\:
\xi_p\,\left(\eta_l - i \gamma_l \right)\;-
\right.
\right.
\nonumber
\end{eqnarray}

\begin{eqnarray}
-\;(k+1)\:
\Omega\left(2\omega_0\,+\,\nu_p,\, \vec{u}_p\right)\:
\left(\eta_p + i\, \gamma_p \right)\, \xi_l \;-
\nonumber
\end{eqnarray}

\begin{eqnarray}
\left.
-\;k\:
\Omega\left(\omega_0\,+\,\nu_p,\, \vec{u}_p\right)\:
\xi_p\,\left(\eta_l + i\, \gamma_l \right) \;-
\;k\:
\Omega\left(\nu_p,\, \vec{u}_p\right)\:
\left(\eta_p + i\, \gamma_p \right)\,\xi_l
\right\} \;+
\nonumber
\end{eqnarray}

\begin{eqnarray}
\left.
+\;
\sum
\limits_{\stackrel{|\nu_l-\nu_p+\omega_0|+}{+|\vec{v}_l-\vec{v}_p|=0}}
\left\{
m\:
\Omega^{*}\left(\nu_p,\, \vec{u}_p\right)\:
\left(\eta_p^{*} - i \gamma_p^{*} \right)\,\xi_l\;-\;
k\:
\Omega^{*}\left(\omega_0\,+\,\nu_p,\, \vec{u}_p\right)\:
\xi_p^{*}\,\left(\eta_l + i \gamma_l \right)
\right\}
\right)
\nonumber
\end{eqnarray}

\begin{eqnarray}
\tilde{c}_2(m,\,k)\;=\; -\frac{m (m-1)}{4}
\sum
\limits_{\stackrel{|\nu_p+\nu_l+2\omega_0|+}{+|\vec{v}_p+\vec{v}_l|=0}}
\Omega\left(\omega_0\,+\,\nu_p,\, \vec{u}_p\right)\:
\xi_p\,\xi_l
\nonumber
\end{eqnarray}

\begin{eqnarray}
\tilde{c}_3(m,\,k)\;=\; \frac{m(m-1)}{4}
\sum
\limits_{\stackrel{|\nu_p+\nu_l+3\omega_0|+}{+|\vec{v}_p+\vec{v}_l|=0}}
\left\{
\Omega\left(\omega_0\,+\,\nu_p,\, \vec{u}_p\right)\:
\xi_p\,\left(\eta_l + i \gamma_l \right)\;+
\right.
\nonumber
\end{eqnarray}

\begin{eqnarray}
\left.
+\;
\Omega\left(2\omega_0\,+\,\nu_p,\, \vec{u}_p\right)\:
\left(\eta_p + i \gamma_p \right)\,\xi_l
\right\}
\nonumber
\end{eqnarray}

\begin{eqnarray}
\tilde{c}_4(m,\,k)\;=\; -\frac{m(m-1)}{4}
\sum
\limits_{\stackrel{|\nu_p+\nu_l+4\omega_0|+}{+|\vec{v}_p+\vec{v}_l|=0}}
\Omega\left(2\omega_0\,+\,\nu_p,\, \vec{u}_p\right)\:
\left(\eta_p + i \gamma_p \right)\,
\left(\eta_l + i \gamma_l \right)
\nonumber
\end{eqnarray}

\begin{eqnarray}
\tilde{c}_5(m,\,k)\;=\; \frac{m}{4}
\left(
\sum
\limits_{\stackrel{|\nu_p+\nu_l+2\omega_0|+}{+|\vec{v}_p+\vec{v}_l|=0}}
\left\{
k\:
\Omega\left(\nu_p,\, \vec{u}_p\right)\:
\left(\eta_p + i \gamma_p \right)\,
\left(\eta_l + i \gamma_l \right)\;+
\right.
\right.
\nonumber
\end{eqnarray}

\begin{eqnarray}
+\;(k+1)\:
\Omega\left(2\omega_0\,+\,\nu_p,\, \vec{u}_p\right)\:
\left(\eta_p + i \gamma_p \right)\,
\left(\eta_l + i \gamma_l \right)\;-
\nonumber
\end{eqnarray}

\begin{eqnarray}
\left.
-\;(m-1)\:
\Omega\left(2\omega_0\,+\,\nu_p,\, \vec{u}_p\right)\:
\left(\eta_p + i \gamma_p \right)\,
\left(\eta_l - i \gamma_l \right)\right\}\;-
\nonumber
\end{eqnarray}

\begin{eqnarray}
\left.
-\;
\sum
\limits_{\stackrel{|\nu_l-\nu_p+2\omega_0|+}{+|\vec{v}_l-\vec{v}_p|=0}}
m\:
\Omega^{*}\left(\nu_p,\, \vec{u}_p\right)\:
\left(\eta_p^{*} - i \gamma_p^{*} \right)\,
\left(\eta_l + i \gamma_l \right)
\right)
\nonumber
\end{eqnarray}

\begin{eqnarray}
\tilde{c}_6(m,\,k)\;=\; \frac{m k}{2}
\sum \limits_{p = -q}^{q}
\frac{|\vec{u}_p|^2}
{|\vec{u}_p|^4 \:+\:\left(\nu_p\:+\:\omega_0 \right)^2}\;
|\xi_p|^2
\nonumber
\end{eqnarray}

\begin{eqnarray}
\tilde{c}_7(m,\,k)\;=\; -\frac{m+k}{2}\,\alpha\;+\;
\frac{m k}{2}
\sum \limits_{p = -q}^{q}
\frac{|\vec{u}_p|^2}
{|\vec{u}_p|^4\:+\:\nu_p^2}\;
|\eta_p\:+\:i\gamma_p|^2 \;+
\nonumber
\end{eqnarray}
\begin{eqnarray}
+\;\frac{m^2+k^2}{4}\,
\sum \limits_{p = -q}^{q}
\frac{|\vec{u}_p|^2}
{|\vec{u}_p|^4\:+\:\nu_p^2}\;
\left(|\gamma_p|^2\:-\:|\eta_p|^2\right)
\;+
\nonumber
\end{eqnarray}
\begin{eqnarray}
+\;\frac{m+2mk+k}{4}\,
\sum \limits_{p = -q}^{q}
\frac{|\vec{u}_p|^2} 
{|\vec{u}_p|^4\:+\:\left(\nu_p\:+\:2\,\omega_0\right)^2}\;
|\eta_p\:+\:i\gamma_p|^2 \;+
\nonumber
\end{eqnarray}
\begin{eqnarray}
+\;i\,\frac{m^2-k^2}{4}\,
\sum \limits_{p = -q}^{q}
\frac{|\vec{u}_p|^2}
{|\vec{u}_p|^4\:+\:\nu_p^2}\; 
\left(\eta_p\,\gamma_p^*\:+\:\eta_p^*\gamma_p\right)
\;+
\nonumber
\end{eqnarray}
\begin{eqnarray}
+\;i\,\frac{m-k}{4}\,
\sum \limits_{p = -q}^{q}
\frac{\nu_p \:+\: 2\,\omega_0} 
{|\vec{u}_p|^4\:+\:\left(\nu_p\:+\:2\,\omega_0\right)^2}\;
|\eta_p\:+\:i\gamma_p|^2 
\nonumber
\end{eqnarray}

{\bf Theorem E:$\;$} 
{\sl 
For any initial points $\;x_0$, $\;z_0$, $\;\vec{y}_0\;$,
for any nonnegative integer $\;N\;$ and for any positive $\;L$
\begin{eqnarray}
\lim_{\varepsilon \rightarrow 0} 
\hspace{0.3cm}
\max_{0 \leq t \leq L / \varepsilon^2}
\hspace{0.3cm}
\left|
\left<
\tilde{\cal M}_N^{-1} (\varepsilon^2 t)
\:\vec{\cal V}(I_{m,\:k}(t);\,N)
 \;-\;  
\vec{\cal V}(I_{m,\:k}(0);\,N)  \right> \right| = 0
\nonumber 
\end{eqnarray} 
where 
the matrix $\;\tilde{\cal M}_N(\tau)\;$ is the fundamental
matrix 
solution of the system of linear
ordinary differential equations with constant coefficients
\begin{eqnarray}
\frac{d}{d \tau}\:\vec{\cal V}(a_{m,\:k};\,N)\;=\;
\tilde{\cal K}_{N} \; \vec{\cal V}(a_{m,\:k};\,N)
\nonumber
\end{eqnarray}
generated with the help of the rule
(\ref{rule_r_breve}) in which we use
$\:\tilde{c}_l(m,\,k)\:$ instead of
$\:\breve{c}_l(m,\,k)$.
}

\section{Proof of the Theorems}

The purpose of this section is to give a 
detailed proof of the theorems.
\subsection{Proof of the Theorem A}

\hspace*{0.5cm}
{\bf 1.} From the fact that all eigenvalues of the matrix $A$ 
have negative real parts it follows that
there exists a quadratic form $\:v(\:\vec{y}\:)\:$ 
satisfying the conditions 
\begin{eqnarray} 
C_1 \:|\vec{y}|^2 \; \leq \; v(\:\vec{y}\:) \;\leq \;
C_2 \:|\vec{y}|^2
\nonumber
\end{eqnarray}
\begin{eqnarray} 
A \vec{y} \cdot \mbox{grad}_{\:\vec{y}} \: v(\:\vec{y}\:) \; \leq \;
-C_3 \:|\vec{y}|^2
\nonumber
\end{eqnarray}
Here and below $C_i$ are some positive constants the
exact values of which are unimportant for us.

{\bf 2.}
Let $\:\hat{L}\:$ be the generating differential operator of the 
$n$-dimensional Markovian diffusion process $\:\vec{y}\:$ i.e.
\begin{eqnarray}
\hat{L} \; = \; 
\frac{\partial}{\partial t} \; + \; 
A \vec{y} \cdot \mbox{grad}_{\:\vec{y}}
\; + \; \frac{1}{2} \: B B^{\top} \: 
\mbox{grad}_{\:\vec{y}} \cdot \mbox{grad}_{\:\vec{y}} \; 
\label{gdo1}
\end{eqnarray}

{\bf 3.}
Introduce the function 
$\;w(\:\vec{y}\:) \;=\; \exp(\psi \:v(\:\vec{y}\:))\;$
for wich
\begin{eqnarray} 
\mbox{grad}_{\:\vec{y}}\:w \;=\;
\psi \: w \cdot
\mbox{grad}_{\:\vec{y}}\:v
\nonumber
\end{eqnarray}
\begin{eqnarray} 
\frac{\partial^2 w}{\partial y_m \partial y_k} \;=\;
\psi 
\left(
\frac{\partial^2 v}{\partial y_m \partial y_k} \:+\:
\psi\:
\frac{\partial v}{\partial y_m} \cdot
\frac{\partial v}{\partial y_k} 
\right) w
\nonumber
\end{eqnarray}
and hence
\begin{eqnarray} 
\hat{L} w = 
\psi 
\left(
A \vec{y} \cdot \mbox{grad}_{\:\vec{y}}\,  v
\:+\:
\frac{1}{2} \sum \limits_{m,\:k \:=\:1}^{n}
\left(B B^{\top}\right)_{m,k}
\left(
\frac{\partial^2 v}{\partial y_m \partial y_k} \:+\:
\psi\:
\frac{\partial v}{\partial y_m} \cdot
\frac{\partial v}{\partial y_k} 
\right)
\right) w
\nonumber
\end{eqnarray}

From this it follows that there exist
constants $C_4$ and $C_5$ independent of the
value of $\psi$ such that 
\begin{eqnarray} 
\hat{L}\: w \;\leq\; \psi\:\left(
-C_3\: |\vec{y}|^2 \:+\:\frac{1}{2} C_4
\:+\:\psi\: C_5 \: |\vec{y}|^2
\right) w
\nonumber
\end{eqnarray}
Taking now $\psi = C_3 / 2 C_5$ we get
\begin{eqnarray} 
\hat{L} \:w \;\leq \;
\frac{\psi}{2} \left(-C_3\: |\vec{y}|^2 \:+\: C_4 \right) w
\label{b1}
\end{eqnarray}

{\bf 4.}
Define the constant $\chi$  as the  maximum of the right side in
inequality (\ref{b1}) with respect to the variables $\vec y$ 
\begin{eqnarray} 
\chi\; =\; \max_{\vec{y} \in R^n}\; \left(
\frac{\psi}{2} \left(-C_3\: |\vec{y}|^2 + C_4 \right) w
\right)
\label{b2}
\end{eqnarray}
Obviously one has
\begin{eqnarray} 
0 \;\leq\; \chi \;\leq\; \frac{\psi}{2}\: C_4 \:
\exp\left(\psi \frac{C_2\: C_4}{C_3}\right)
\nonumber
\end{eqnarray}

From (\ref{b1}) and (\ref{b2}) it follows, that the function
\begin{eqnarray} 
\hat{w} \;=\; w \:+\: \chi\:\left(\frac{L}{\varepsilon^2} \;-\; t \right)
\label{qq1}
\end{eqnarray}
will satisfy the inequality
\begin{eqnarray} 
\hat{L} \,\hat{w} \;\leq \;0
\label{qq2}
\end{eqnarray}

{\bf 5.}
Let 
$\;\tilde{s}^{\varepsilon}_t \:=\:  
s^{\varepsilon}_t \:\wedge \:\frac{L}{\varepsilon^2}\;$.
From (\ref{qq1}) and (\ref{qq2}) immediately follows that the
stochastic process
$\;\hat{w}\left(\:\tilde{s}^{\varepsilon}_t \:\right)\;$ 
is a nonnegative supermartingale, and hence
\begin{eqnarray} 
P \left( \tau_{\varepsilon} \:<\: 
\frac{L}{\varepsilon^2}\right) \;\leq\; 
P \left(\sup_{t\:\geq\:0}\left|\vec{y}\left( \tilde{s}_{t}^{\varepsilon}
\right)\right| \geq c(\varepsilon)\right) \leq 
\nonumber
\end{eqnarray}

\begin{eqnarray} 
P \left(\sup_{t \geq 0}\hat{w}\left( \tilde{s}_{t}^{\varepsilon}
\right) \geq \exp(\psi C_1 c^2(\varepsilon))\right) \leq 
\left(\exp(\psi v(\vec{y}_0)) + \chi \frac{L}{\varepsilon^2}\right) 
\exp(-\psi C_1 c^2(\varepsilon)) 
\nonumber
\end{eqnarray}
The first two inequalities in the sequence shown above are almost obvious,
and the last one follows from the property of the
stochastic process
$\;\hat{w}\left(\tilde{s}^{\varepsilon}_t \right)\;$ to be a nonnegative
supermartingale. For finishing the proof
take $a = \max(\chi,\: C_2 \psi)$ and $\;b =  C_1 \psi\;$.

\subsection{Proof of the Theorem B}

\hspace{0.5cm}
{\bf 1.} The joint solution of the systems (\ref{m02}), (\ref{ss1})
is a Markovian diffusion process in the ($n+2$)-dimensional
Euclidean space. Let $\:L\:$ be the generating differential operator 
of this stochastic process. Separating the orders according to
$\:\varepsilon\:$ we can represent $\:L\:$ in the form
\begin{eqnarray}
L \;=\; L_0 \:+\: 
\varepsilon   \: L_{\varepsilon} \:+\:
\varepsilon^2 \: L_{\varepsilon^2}
\label{m03}
\end{eqnarray}
where the differential operators $\:L_0\:$, $\:L_{\varepsilon}\:$,
$\:L_{\varepsilon^2}\:$ are defined as follows
\begin{eqnarray}
L_0 = \omega_0 \: 
\left( 
z \: \frac{\partial}{\partial x} - 
x \: \frac{\partial}{\partial z}
\right)
+ \hat{L}, 
\hspace{0.6cm}
L_{\varepsilon} = 
\left(
\xi - \gamma \: z - \eta \: x
\right) 
\frac{\partial}{\partial z}, 
\hspace{0.6cm}
L_{\varepsilon^2}  = -\alpha \: z \:
\frac{\partial}{\partial z}
\nonumber
\end{eqnarray}
and $\:\hat{L}\:$ is the generating differential operator of the 
$n$-dimensional Markovian diffusion process $\:\vec{y}\:$ 
given by (\ref{gdo1}).

{\bf 2.}
Now we wish to show that there exist functions 
$\:u_{m,\: k}^{\varepsilon}\:$ satisfying 
\begin{eqnarray}
L_0 \:u_{m,\: k}^{\varepsilon} \;= \; 
- L_{\varepsilon} \: I_{m,\: k} 
\label{star}
\end{eqnarray}
Representing the operator $\: L_{\varepsilon} \:$ in the form
\begin{eqnarray}
L_{\varepsilon} \;=\; 
\left(
\rule[0.25cm]{0cm}{0.25cm}
\xi \;-\; (\eta \:+\: i \gamma) \:\exp(i \:\omega_0\: t)\: I_{0,\:1}
\;-\; (\eta \:-\: i \gamma) \:\exp(-i\: \omega_0 \:t)\: I_{1,\: 0}
\right) \cdot
\frac{\partial}{\partial z} 
\nonumber
\end{eqnarray}
and calculating
\footnote{Starting from this point it is convenient to
extend the definition of the function $\:I_{p,\:q}\:$
to negative indices 
assuming that if $\:p < 0\:$ or $\:q < 0\:$
then $\:I_{p,\:q}\:\equiv\:0\:$.
In general, after this extension one has to be
careful with respect to the application of the property {\bf b},
but we have not to worry about it, because the only
 source of lowering indices   in this paper
is differentiation and hence if the function
$\:I_{p,\:q}\:$ with negative index will appear we
shall have automatically zero multiplyer in front of it.  
}
\begin{eqnarray}
\frac{\partial I_{m, \: k}}{\partial z} \: = \:
i \frac{m}{2}\: \exp( i \: \omega_0 \: t) \; I_{m-1,\: k} \; - \;
i \frac{k}{2}\: \exp(-i \: \omega_0 \: t) \; I_{m, k-1} 
\label{re4}
\end{eqnarray}
and taking into account property {\bf b} 
we have
\begin{eqnarray}
L_{\varepsilon} \; I_{m, \: k} \;=\; 
\frac{i}{2} 
\left(\rule[0.25cm]{0cm}{0.25cm}
m\:\xi\:\exp(i\:\omega_0\:t)\;I_{m-1,\: k}\;-\;
k \: \xi \: \exp(-i \: \omega_0 \: t) \; I_{m, \: k-1} \; + 
\right. 
\nonumber
\end{eqnarray}
\begin{eqnarray}
\left[\: k \: (\eta \: + \: i \gamma) 
\;-\; m \:(\eta \:-\: i \gamma)\right]\; I_{m,\: k}
\;+ 
\nonumber
\end{eqnarray}
\begin{eqnarray}
\left.
k \: (\eta \:-\: i \gamma) \: \exp(-i\: 2\: \omega_0\: t)\;
I_{m+1,\:k-1}\;- \; 
m \:(\eta \:+ \:i \gamma) \: \exp( i\: 2\: \omega_0\: t) 
\;I_{m-1,\: k+1} 
\rule[0.25cm]{0cm}{0.25cm} \right)
\nonumber
\end{eqnarray}
Looking for the $\;u_{m,\:k}^{\varepsilon}\;$ in analogous form
\begin{eqnarray}
u_{m, k}^{\varepsilon} \:=\: 
\frac{i}{2} \left(\rule[0.25cm]{0cm}{0.25cm} 
- m \:a_1 \:  \exp( i\: \omega_0 \:t)\; I_{m-1,\: k}
\;+\; k\: a_1^* \:\exp(-i \:\omega_0\: t)\; I_{m,\: k-1} \;+ \right.
\nonumber
\end{eqnarray}
\begin{eqnarray}
(m \:a_2^* \;-\; k\: a_2)\; I_{m,\: k} \; +
\nonumber
\end{eqnarray}
\begin{eqnarray}
\left.
m\: a_3 \:  \exp( i\: 2\: \omega_0\: t)\; I_{m-1,\: k+1}
\;-\; k\: a_3^*\: \exp(-i\: 2\: \omega_0\: t)\; I_{m+1,\: k-1}
\rule[0.25cm]{0cm}{0.25cm} 
\right)
\nonumber
\end{eqnarray}
we get the system defining the unknown $\:a_l\:$ 
\begin{eqnarray}
\left\{
\begin{array}{l}
\hat{L} \: a_1 \;+\; i \omega_0 \: a_1 \;=\; \xi \\
\\
\hat{L} \: a_2 \;=\; \eta \:+\: i \gamma \\
\\
\hat{L} \: a_3 \;+\; i \:2\: \omega_0\: a_3 \;=\;
 \eta \:+\: i \gamma 
\end{array}
\right.
\label{als1}
\end{eqnarray}
Choosing $\;a_l\;$ in the form 
$\;a_l \;=\; \vec{a}_l(t) \cdot \vec{y}\;$, where
\begin{eqnarray}
\vec{a}_l(t) \;=\; 
\sum \limits_{p = -\infty}^{+\infty} \;
\vec{a}_{l,\:p} \: \exp(i \:\nu_p \:t)
\label{vk}
\end{eqnarray}
and taking into account that
\begin{eqnarray}
\hat{L} \:a_l \;=\; 
\left(\frac{d \vec{a}_l}{d t} \;+\; 
A^{\top } \:\vec{a}_l \right) \cdot \vec{y}
\nonumber
\end{eqnarray}
we reduce the system (\ref{als1}) to a system of
algebraic equations for the Fourier coefficients
\begin{eqnarray}
\left\{
\begin{array}{l}
\Lambda^{\top}(\omega_0\:+\:\nu_p)\:\vec{a}_{1,\:p} 
\; = \; \vec{h}_p \\ 
\\
\Lambda^{\top}(\nu_p) \: \vec{a}_{2,\:p} \; = \; 
\vec{b}_p \: + \: i \vec{d}_p \\ 
\\
\Lambda^{\top}(2\: \omega_0\:+\:\nu_p) \: \vec{a}_{3,\:p} \; = \; 
\vec{b}_p \: + \: i \vec{d}_p \\ 
\end{array}
\right.
\label{als1_2}
\end{eqnarray}
where we have used the notation
\begin{eqnarray}
\Lambda(\omega) \;=\; A \:+\: i\,\omega \,I.
\nonumber
\end{eqnarray}
So among the characteristic roots of $\:A\:$ we have no
purely imaginary or zero values the matrix $\:\Lambda(\omega)\:$ is
invertiable
for an arbitrary real $\:\omega\:$ and hence the system (\ref{als1_2})
has a unique solution, which can be expressed as follows
\begin{eqnarray}
\vec{a}_{1,\:p} 
\; = \; 
\Lambda^{-\top}(\omega_0\:+\:\nu_p)
\;\vec{h}_p 
\; = \; 
\frac{\Lambda^{*}(\omega_0\:+\:\nu_p)}
{A^{\top}\:A^{\top} \:+ \:(\omega_0\:+\:\nu_p)^2 I}
\;\vec{h}_p
\nonumber
\end{eqnarray}
\begin{eqnarray}
\vec{a}_{2,\:p} 
\; = \; 
\Lambda^{-\top}(\nu_p)
\;\left(\vec{b}_p \:+\:i\vec{d}_p\right) 
\; = \; 
\frac{\Lambda^{*}(\nu_p)}
{A^{\top}\:A^{\top} \:+ \:\nu_p^2 I}
\;\left(\vec{b}_p \:+\:i\vec{d}_p\right) 
\nonumber
\end{eqnarray}
\begin{eqnarray}
\vec{a}_{3,\:p} 
\; = \; 
\Lambda^{-\top}(2\omega_0\:+\:\nu_p)
\;\left(\vec{b}_p \:+\:i\vec{d}_p\right) 
\; = \; 
\frac{\Lambda^{*}(2\omega_0\:+\:\nu_p)}
{A^{\top}\:A^{\top} \:+ \:(2\omega_0\:+\:\nu_p)^2 I}
\;\left(\vec{b}_p \:+\:i\vec{d}_p\right)
\nonumber
\end{eqnarray}
Using the estimate
\begin{eqnarray}
\left|
{\Lambda}^{-\top}(\omega)
\right| \; \leq \; \frac{1}{\delta_s^2}
\nonumber
\end{eqnarray}
which is valid for an arbitrary real $\;\omega\;$ we get
for  $\;\vec{a}_{l,\:p}\;$ 
\begin{eqnarray}
\left| \vec{a}_{l,\:p} \right| \: \leq \: 
\frac{1}{\delta_s^2}
\: \left(
\left| \vec{h}_{p} \right| +  
\left| \vec{b}_{p} \right| +  
\left| \vec{d}_{p} \right|\right)
\nonumber
\end{eqnarray}
which together with (\ref{cond1}) guarantees the 
absolute
convergence and the possibility of differentiating the 
series (\ref{vk}) term by term.

{\bf 3.}
Calculating $\;L_{\varepsilon^2}\: I_{m,\:k}\;$ and 
$\;L_{\varepsilon}\: u_{m,\:k}^{\varepsilon}\;$ we get
\begin{eqnarray}
\hspace*{-0.2cm}
L_{\varepsilon^2} \: I_{m,\:k} \; + \;
L_{\varepsilon} \: u_{m,\:k}^{\varepsilon} \;=
\nonumber
\end{eqnarray}
\vspace*{-0.8cm}
\begin{eqnarray}
\begin{array}{llll}
c_2(m,\:k) \: I_{m-2,\:k} & + &
c_2^*(k,\:m) \: I_{m,\:k-2} & + \\
\\
c_1(m,\:k) \: I_{m-1,\:k} & + &
c_1^*(k,\:m) \: I_{m,\:k-1} & + \\
\\
c_3(m,\:k) \: I_{m-2,\:k+1} & + &
c_3^*(k,\:m) \: I_{m+1,\:k-2} & + \\
\\
c_4(m,\:k) \: I_{m-2,\:k+2} & + &
c_4^*(k,\:m) \: I_{m+2,\:k-2} & + \\
\\
c_5(m,\:k) \: I_{m-1,\:k+1}  & + &
c_5^*(k,\:m) \: I_{m+1,\:k-1} & + \\
\\
c_6(m,\:k) \: I_{m-1,\:k-1}  & + &
c_7(m,\:k) \: I_{m,\:k} &
\end{array}
\label{ccil}
\end{eqnarray}
where the functions $\;c_l(m,\:k)\;$ are given by
the following expressions
\begin{eqnarray}
\begin{array}{lll}
c_1(m, \:k) & = &
\hspace{0.3cm}
\frac{m}{4}\: 
\left[\:
\rule[0.2cm]{0cm}{0.2cm}
(\:k\: a_2 \:-\: m \:a_2^*\: +\: (k+1) \: a_3 \:) \:\xi 
\right. \:+\\
\\
& &
\hspace{0.3cm}
\left.
k\: (a_1 \:+\: a_1^*) (\eta \:+\: i \gamma)
\;-\;(m-1) \:a_1\:(\eta \:-\: i \gamma)
\rule[0.2cm]{0cm}{0.2cm}
\right] \:
\exp(i\: \omega_0\: t)\\
\\
c_2(m, \:k) & = &
\hspace{0.3cm}
\frac{m}{4} \:
\left[
\rule[0.2cm]{0cm}{0.2cm}
\: (m-1) \: a_1 \: \xi 
\right] \: \exp(i\: 2\:\omega_0\: t)\\
\\
c_3(m, \:k) & = &
-\frac{m}{4} \:
\left[
\rule[0.2cm]{0cm}{0.2cm}
\: (m-1) \: a_1 \: (\eta \: + \: i\gamma) \;+\;
(m-1) \: a_3 \: \xi \:\right] \:
\exp(i\: 3\:\omega_0\: t)\\
\\
c_4(m, \:k) & = &
\hspace{0.3cm}
\frac{m}{4}\:
\left[
\rule[0.2cm]{0cm}{0.2cm}
\:(m-1)\:a_3\:(\eta\:+\:i\gamma)\:\right]
\:\exp(i\: 4\:\omega_0\: t)\\
\\
c_5(m, \:k) & = &
\hspace{0.3cm}
\frac{m}{4} 
\left[
\rule[0.2cm]{0cm}{0.2cm}
2\alpha \;-\;
(\:k\: a_2 \:-\: m \:a_2^*\: +\: (k+1) \:a_3) \:
(\eta \:+\:i\gamma)
\right.\:+\\
\\
& &
\hspace{0.3cm}
\left.
(m-1) \:a_3\:(\eta \:-\: i \gamma)
\rule[0.2cm]{0cm}{0.2cm}
\right] \:
\exp(i\: 2\:\omega_0\: t)\\
\\
c_6(m, \:k) & = &
-\frac{m}{4}\:\left[
\rule[0.2cm]{0cm}{0.2cm}
k\:(a_1\:+\:a_1^*)\: \xi\right] \\
\\
c_7(m, \:k) & = &
-\frac{m}{4}
\left[
\rule[0.2cm]{0cm}{0.2cm}
2\alpha \;+\; (k\:a_2\:-\:m\:a_2^* \:+\: (k+1)\:a_3)
(\eta\:-\:i\gamma)\right]\;- \\
\\
& &
\hspace{0.3cm}
\frac{k}{4}
\left[
\rule[0.2cm]{0cm}{0.2cm}
2\alpha \;+\; (m\:a_2^*\:-\:k\:a_2 \:+\: (m+1)\:a_3^*)
(\eta\:+\:i\gamma)\right] 
\end{array}
\nonumber
\end{eqnarray}
Note that 
$\;c_6(m,\:k)\:=\:c_6^*(k,\:m)\;$ and
$\;c_7(m,\:k)\:=\:c_7^*(k,\:m)\:$.

{\bf 4.}
Introduce a $(n \times n)$ matrix 
$\:K(\:\vec{y}\:)\:=\: \vec{y} \cdot \vec{y}^{\,\top}\:$
with the elements $\:k_{ij}\:=\:y_i\:y_j\:$. It is easy
to check, that this matrix satisfies the equation
\begin{eqnarray} 
\hat{L} \: K \;=\; A K \:+\: K A^{\top} \:+\: B \: B^{\top}
\nonumber
\end{eqnarray}
The usefulness of this matrix for the following is connected 
with the fact that
for arbitrary complex vectors $\;\vec{a}\;$ and $\;\vec{c}\;$
\begin{eqnarray} 
(\vec{a}\cdot\vec{y}\:) \cdot (\vec{c}\cdot\vec{y}\:)
\;=\;K\: \vec{a} \cdot \vec{c}^{\;*}
\;=\;K\: \vec{c} \cdot \vec{a}^{\;*}
\label{mju}
\end{eqnarray}

{\bf 5.}
Define the $(n \times n)$ matrix-function
$\:P_{\omega} = P_{\omega}(\,\vec y\,)\:$ with the help of 
the integral
\begin{eqnarray} 
P_{\omega} \;=\; - \int \limits_{0}^{\infty}
\exp(i \omega \tau) \exp(A \tau) 
K(\vec{y})
\exp(A^{\top} \tau)\: d \tau
\label{i1}
\end{eqnarray}
This integral converges because all the characteristic roots of
$\:A\:$ have negative real parts. Introduce the new integration variable
$\;\tau^{\prime} = \tau + t\;$, where $\:t\:$ is some parameter. 
Then (\ref{i1}) becomes
\begin{eqnarray} 
P_{\omega} \;=\; - \int \limits_{t}^{\infty}
\exp(i \omega (\tau^{\prime} - t)) \exp(A (\tau^{\prime} - t)) 
K(\vec{y})
\exp(A^{\top} (\tau^{\prime} - t)) \:d \tau^{\prime}
\label{i2}
\end{eqnarray}
Differentiating (\ref{i2}) with respect to $\:t\:$ and using that 
due to (\ref{i1}) $\:P_{\omega}\:$ does not depend on $\:t\:$, we obtain
\begin{eqnarray} 
\frac{d P_{\omega}}{d t} \;=\; 
K \:-\: A P_{\omega} \:-\: P_{\omega} A^{\top} \:-\: 
i \omega P_{\omega} \;=\; 0
\label{30}
\end{eqnarray}
Calculating $\;\hat{L} \: P_{\omega}\;$ and taking into account 
(\ref{30}) we get
\begin{eqnarray} 
\hat{L} \: P_{\omega} \;=\; A P_{\omega} \:+\: P_{\omega} A^{\top} 
\:-\:C(\omega)\;=\; 
- i \omega P_{\omega} \:+\: K \:-\: C(\omega)
\label{31}
\end{eqnarray}
where we have introduced the notation
\begin{eqnarray} 
C(\omega) \;=\; \int \limits_{0}^{\infty}
\exp(i \omega \tau)
\exp(A \tau) B B^{\top} \exp(A^{\top} \tau) \:d \tau,
\hspace*{1.0cm}
C(0) = D
\nonumber
\end{eqnarray}
For the following let us rewrite (\ref{31}) in the form
\begin{eqnarray}
\hat{L} \:P_{\omega} \:+\: i \omega P_{\omega} 
\;=\; K \:-\: C(\omega)
\label{pj1}
\end{eqnarray}
Note that for some positive constant $C_1$ the norms of
the matrices $P_{\omega}$ and $C(\omega)$ 
can be estimated 
uniformly with respect to real $\omega$
as follows (using \ref{SS11})
\begin{eqnarray}
\left|\:P_{\omega}\:\right| 
\;\leq\; \frac{C_1}{\delta_s^2}\: |\vec{y}|^2,
\hspace*{1.0cm}
\left|\:C(\omega)\:\right| 
\;\leq\; \frac{C_1}{\delta_s^2}
\label{nqpw}
\end{eqnarray}

{\bf 6.}
Now we wish to show that there exist functions 
$\:g_l(m,\: k)\:$
\footnote{Of course,
$\:g_l(m,\: k)\:$
like
$\:c_l(m,\: k)\:$
are also functions of
$t$ and $\vec{y}$}
satisfying 
\begin{eqnarray} 
\hat{L}\:g_l(m,\: k)\;+\;
c_l(m,\: k)\;=\;\bar{c}_l(m,\: k),
\hspace*{1.0cm} l\;=\;1,\ldots,7
\nonumber
\end{eqnarray}
and these functions have continuous first and
continuous first and second derivatives with
respect to the variables $\:t\:$ and $\:\vec{y}\:$
respectively, and these functions together with
the above derivative
are bounded 
with respect to the 
 variable $\,t\,$ for fixed
values of $\:m,\,k,\,\vec y\,$.

We shall show this for $\:l\:=\:2\:$ and the rest can be
done by analogy.

Due to (\ref{mju}) and the reality of the vector $\:\vec{h}\:$
(that is $\:\vec{h}\:=\:\vec{h}^{\;*}\:$) we have
\begin{eqnarray} 
c_2(m, \:k) \; = \;
\frac{m(m-1)}{4} \: a_1 \: \xi 
\: \exp(i\: 2\:\omega_0\: t) \;=\;
\nonumber
\end{eqnarray}
\begin{eqnarray} 
\frac{m(m-1)}{4}  
\left(
\rule[0.2cm]{0cm}{0.2cm}
K \: \vec{a}_1 \cdot \vec{h}^{\;*}
\right) \exp(i\: 2\:\omega_0\: t) \;=\;
\frac{m(m-1)}{4} 
\left(
\rule[0.2cm]{0cm}{0.2cm}
K \: \vec{a}_1 \cdot \vec{h}
\right) \exp(i\: 2\:\omega_0\: t)
\nonumber
\end{eqnarray}
Substituting in the last expression the
Fourier series of the vectors $\:\vec{a}_1\:$ and
$\:\vec{h}\:$ we transform $\:c_2(m,\:k)\:$ into 
the form
\begin{eqnarray} 
c_2(m, \:k) \; = \;
\frac{m(m-1)}{4} \: 
\sum \limits_{p,\:l\;=\;-\infty}^{\infty}
\left(K\:\vec{a}_{1,\:p} \cdot \vec{h}_l\right) 
\: \exp(i(\nu_p-\nu_l+2\omega_0) t) \;=
\nonumber
\end{eqnarray}
\begin{eqnarray} 
\frac{m(m-1)}{4} \left\{ 
\sum \limits_{\nu_l-\nu_p\:=\:2\omega_0}
K \vec{a}_{1,\:p} \cdot \vec{h}_l +
\sum \limits_{\nu_l-\nu_p\:\neq\:2\omega_0}
\left(K \vec{a}_{1,\:p} \cdot \vec{h}_l\right)
\exp(i(\nu_p-\nu_l+2\omega_0) t) 
\right\} 
\nonumber
\end{eqnarray}
For $\:\omega \neq 0\;$ introduce the matrix
\begin{eqnarray} 
Q(\omega) \; = \; \frac{i}{\omega} C(\omega) \:-\:P_{\omega}
\nonumber
\end{eqnarray}
and denote $P \;=\; -P_0$. 
Taking into account (\ref{pj1}) we have 
\begin{eqnarray} 
\hat{L}\:\left(Q(\omega)\exp(i \omega t)\right) \; = \;
-\:K \:\exp(i \omega t) 
\hspace*{1.0cm}
\mbox{and}
\hspace*{1.0cm}
\hat{L}\:P \;=\; D\:-\:K
\nonumber
\end{eqnarray}
Choosing now
\begin{eqnarray} 
g_2(m,\:k)\;=\;\frac{m(m-1)}{4} \left\{ 
\sum \limits_{\nu_l-\nu_p\:=\:2\omega_0}
P \vec{a}_{1,\:p} \cdot \vec{h}_l \;+\;
\right.
\nonumber
\end{eqnarray}
\begin{eqnarray} 
\left.
\sum \limits_{\nu_l-\nu_p\:\neq\:2\omega_0}
\left(Q(\nu_p-\nu_l+2\omega_0) \vec{a}_{1,\:p} \cdot \vec{h}_l\right)
\: \exp(i(\nu_p-\nu_l+2\omega_0) t) 
\right\} 
\nonumber
\end{eqnarray}
we obtain
\begin{eqnarray} 
\hat{L}\:g_2(m,\: k)\;+\;
c_2(m,\: k)\;=\;
\frac{m(m-1)}{4} 
\sum \limits_{\nu_l-\nu_p\:=\:2\omega_0}
D\:\vec{a}_{1,\:p} \cdot \vec{h}_l
\nonumber
\end{eqnarray}
which just coincides with the expression for $\:\bar{c}_2(m,\:k)\:$
if we take into account that
\begin{eqnarray}
D\,\vec{a}_{1,\:p} 
\; = \; 
D\:\frac{\Lambda^{*}(\omega_0\:+\:\nu_p)}
{A^{\top}\:A^{\top} \:+ \:(\omega_0\:+\:\nu_p)^2 I}
\;\vec{h}_p \;=\;
-\Psi^*(\omega_0\:+\:\nu_p) 
\;\vec{h}_p
\label{muy7}
\end{eqnarray}
Due to (\ref{nqpw}) and (\ref{cond2}) we have
\begin{eqnarray}
\max
\left\{\:
\left|\:P\:\right|, \; 
\max_{\nu_l-\nu_p\:\neq\:2\omega_0}
\left|\:Q(\nu_p-\nu_l+2\omega_0)\:\right|\:
\right\} 
\;\leq\; 
\frac{C_1}{\delta_s^2} 
\left(
|\vec{y}|^2 +
\frac{1}{\delta_f^2} 
\right)
\nonumber
\end{eqnarray}
and hence, as it can be easily shown, 
the series defining the function
$\:g_2(m,\:k)\:$ converges absolutely
with  
\begin{eqnarray}
\left|\:
g_2(m,\:k)
\:\right|
\;\leq\; 
\frac{m(m-1)}{4}
\frac{C_1}{\delta_s^2} 
\left(
|\vec{y}|^2 +
\frac{1}{\delta_f^2} 
\right)
\left(
\sum \limits_{p = -\infty}^{\infty}
\left|\vec{a}_{1,\:p}\right|
\right)
\left(
\sum \limits_{p = -\infty}^{\infty}
\left|\vec{h}_{p}\right|
\right)
\label{series1}
\end{eqnarray}

The function $\;g_2(m,\:k)\;$ is a quadratic polynomial
in $\;\vec{y}\;$ and so we need to worry about their
partial derivative with respect to $\;t\;$ only.
Expressing
\begin{eqnarray}
\frac{\partial g_2(m,\:k)}{\partial t}
\;=\;
\frac{m(m-1)}{4} 
\sum \limits_{\mu_{p,\:l}\:\neq\:0}
i\: \mu_{p,\:l} \:
\left(Q(\mu_{p,\:l})\: \vec{a}_{1,\:p} \cdot \vec{h}_l\right)
\:\exp(i \mu_{p,\:l} t) 
\nonumber
\end{eqnarray}
where $\;\mu_{p,\:l}\;=\;\nu_p\:-\:\nu_l\:+\:2\omega_0\;$
and using the very rough estimate for $\;|\mu_{p,\:l}|\;$
\begin{eqnarray}
|\mu_{p,\:l}|\;\leq\;
C_2 (1+|\nu_p|)(1+|\nu_l|),
\hspace{0.7cm}
C_2 \:=\: \max \left\{\:1,\; 2 |\omega_0|\:\right\}
\nonumber
\end{eqnarray}
we get that
the series defining the partial derivative with respect to
the variable $\;t\;$ converges absolutely
with 
\begin{eqnarray}
\left|\:
\frac{\partial g_2(m,\:k)}{\partial t}
\:\right| \:\leq\: 
\frac{m(m-1)}{4}
\frac{C_1}{\delta_s^2} 
\left(
|\vec{y}|^2 +
\frac{1}{\delta_f^2} 
\right) C_2 \cdot
\nonumber
\end{eqnarray}
\begin{eqnarray}
\cdot \left(
\sum \limits_{p = -\infty}^{\infty}
(1+|\nu_p|)
\left|\vec{a}_{1,\:p}\right|
\right)
\left(
\sum \limits_{p = -\infty}^{\infty}
(1+|\nu_p|)
\left|\vec{h}_{p}\right|
\right)
\label{series2}
\end{eqnarray}

Note that expressions similar to (\ref{muy7}) hold also
for  $\;D \vec{a}_{2,\:p}\;$ and $\;D \vec{a}_{3,\:p}$ 
\begin{eqnarray}
D\,\vec{a}_{2,\:p} 
\; = \; -\Psi^*(\nu_p) 
\;\left(\vec{b}_p\:+\:i\vec{d}_p\right)
\nonumber
\end{eqnarray}
\begin{eqnarray}
D\,\vec{a}_{3,\:p} 
\; = \; -\Psi^*(2\omega_0\:+\:\nu_p) 
\;\left(\vec{b}_p\:+\:i\vec{d}_p\right)
\nonumber
\end{eqnarray}

{\bf 7.}
Defining $\;u_{m,\:k}^{\varepsilon^2}\;$ as
\begin{eqnarray}
\begin{array}{llllll}
u_{m,\:k}^{\varepsilon^2} & =
& g_2(m,\:k) \: I_{m-2,\:k} & + &
g_2^*(k,\:m) \: I_{m,\:k-2} & + \\
\\
& & g_1(m,\:k) \: I_{m-1,\:k} & + &
g_1^*(k,\:m) \: I_{m,\:k-1} & + \\
\\
& & g_3(m,\:k) \: I_{m-2,\:k+1} & + &
g_3^*(k,\:m) \: I_{m+1,\:k-2} & + \\
\\
& & g_4(m,\:k) \: I_{m-2,\:k+2} & + &
g_4^*(k,\:m) \: I_{m+2,\:k-2} & + \\
\\
& & g_5(m,\:k) \: I_{m-1,\:k+1}  & + &
g_5^*(k,\:m) \: I_{m+1,\:k-1} & + \\
\\
& & g_6(m,\:k) \: I_{m-1,\:k-1}  & + &
g_7(m,\:k) \: I_{m,\:k} &
\end{array}
\nonumber
\end{eqnarray}
and acting on the function
\begin{eqnarray}
\tilde{I}_{m,\:k} = I_{m,\:k} \:+\:
\varepsilon   \: u_{m,\:k}^{\varepsilon} \:+\:
\varepsilon^2 \: u_{m,\:k}^{\varepsilon^2} 
\label{jj7}
\end{eqnarray}
by means of the operator $L$ we have 
\begin{eqnarray}
L \: \tilde{I}_{m,\:k} \:=\:
\left(
L_0\:+\:
\varepsilon L_{\varepsilon} \:+\:
\varepsilon^2 L_{\varepsilon^2}
\right)\:\tilde{I}_{m,\:k} \:=\:
\nonumber
\end{eqnarray}
\begin{eqnarray}
L_0 \: I_{m,\:k} \:+\:
\varepsilon 
\left(
L_0\:u_{m,\:k}^{\varepsilon}\:+\:
L_{\varepsilon}\:I_{m,\:k}
\right)\:+\:
\nonumber
\end{eqnarray}
\begin{eqnarray}
\varepsilon^2
\left(
L_0\:u_{m,\:k}^{\varepsilon^2}\:+\:
L_{\varepsilon^2}\:I_{m,\:k} \:+\:
L_{\varepsilon}\:u_{m,\:k}^{\varepsilon}
\right) \:+\:
\varepsilon^3 R_{m,\:k}
\label{jj6}
\end{eqnarray}
where for the remainder $R_{m,\:k}$ we have the expression
\begin{eqnarray}
R_{m,\:k}\; =\; L_{\varepsilon^2} \: u_{m,\:k}^{\varepsilon} \:+\:
L_{\varepsilon} \: u_{m,\:k}^{\varepsilon^2} +
\varepsilon L_{\varepsilon^2} \: u_{m,\:k}^{\varepsilon^2}
\label{jj9}
\end{eqnarray}
Due to our construction of the functions
$ \: I_{m,\:k} \:$,
$ \: u_{m,\:k}^{\varepsilon} \:$ and
$ \: u_{m,\:k}^{\varepsilon^2}\:$ from (\ref{jj6}) 
it follows that
\begin{eqnarray}
\begin{array}{llllll}
L \: \tilde{I}_{m,\:k} & =
&\varepsilon^2 \left(
\bar{c}_2(m,\:k) \: I_{m-2,\:k} \right.& + &
\bar{c}_2^{\;*}(k,\:m) \: I_{m,\:k-2} & + \\
\\
& & \hspace*{0.6cm}
\bar{c}_1(m,\:k) \: I_{m-1,\:k} & + &
\bar{c}_1^{\;*}(k,\:m) \: I_{m,\:k-1} & + \\
\\
& & \hspace*{0.6cm}
\bar{c}_3(m,\:k) \: I_{m-2,\:k+1} & + &
\bar{c}_3^{\;*}(k,\:m) \: I_{m+1,\:k-2} & + \\
\\
& & \hspace*{0.6cm}
\bar{c}_4(m,\:k) \: I_{m-2,\:k+2} & + &
\bar{c}_4^{\;*}(k,\:m) \: I_{m+2,\:k-2} & + \\
\\
& & \hspace*{0.6cm}
\bar{c}_5(m,\:k) \: I_{m-1,\:k+1}  & + &
\bar{c}_5^{\;*}(k,\:m) \: I_{m+1,\:k-1} & + \\
\\
& & \hspace*{0.6cm}
\bar{c}_6(m,\:k) \: I_{m-1,\:k-1}  & + &
\left. 
\bar{c}_7(m,\:k) \: I_{m,\:k}\right) &
+\\
\\
& & \varepsilon^3 \; R_{m,\:k}\:
\end{array}
\label{jj76}
\end{eqnarray}

{\bf 8.}
So
\begin{eqnarray}
\left|\:I_{p,\:q}\:\right|\:=\:
\left(\frac{r}{2}\right)^{\frac{p+q}{2}}
\:\leq\: 
1\:+\:
\left(\frac{r}{2}\right)^{\frac{m+k}{2}}\:=\:
1 \:+\:
\left|\:I_{m,\:k}\:\right|
\nonumber
\end{eqnarray}
for $\;p+q \:\leq\: m+k\:$,
then for some positive constant $\:C_3\:$ independent
of  $m$ and $k$ 
the  functions
$\;u_{m,\:k}^{\varepsilon}\;$, 
$\;u_{m,\:k}^{\varepsilon^2}\;$ 
defined above and
$\;R_{m,\:k}\;$ 
can be roughly estimated as follows 
\begin{eqnarray}
\left|\;u_{m,\:k}^{\varepsilon}\;\right|
\; \leq \;
C_3 (m + k) \left| \vec{y} \right|
\left(
1 + 
\left|\;I_{m,\:k}\;\right|
\right)
\label{jo1}
\end{eqnarray}
\begin{eqnarray}
\left|\;u_{m,\:k}^{\varepsilon^2}\;\right|
\; \leq \;
C_3 (m + k)^2 
\left(1 +
\left| \vec{y} \right|^2
\right)
\left(
1 + 
\left|\;I_{m,\:k}\;\right|
\right)
\label{jo2}
\end{eqnarray}
\begin{eqnarray}
\left|\;R_{m,\:k}\;\right|
\; \leq \;
C_3 (m + k)^3 
\left(1 +
\left| \vec{y} \right|^3
\right)
\left(
1 + 
\left|\;I_{m,\:k}\;\right|
\right)
\label{jo3}
\end{eqnarray}

{\bf 9.}
So 
$\;\left|\:I_{m,\:m}\:\right|\:=\: I_{m,\:m}\;$
then for some positive constant $C_4$ independent
of $m$ we have from (\ref{jj76})
\begin{eqnarray}
L\: \tilde{I}_{m,\:m}
\; \leq \;
\varepsilon^2 \: C_4\: m^2\:
\left(
1\:+\:I_{m,\:m}
\right)
\;+\; \varepsilon^3
\left|\;R_{m,\:m}\;\right|
\label{jo3_1}
\end{eqnarray}
Using the estimate (\ref{jo3}) we can rewrite
(\ref{jo3_1}) in the form
\begin{eqnarray}
L\: \tilde{I}_{m,\:m}
\; \leq \;
\varepsilon^2 \: {\cal H}_{m}\:
\left(
1\:+\:I_{m,\:m}
\right)
\label{jo3_2}
\end{eqnarray}
and for 
$\;u_{m,\:m}^{\varepsilon} \:+\:
\varepsilon \:u_{m,\:m}^{\varepsilon^2}\;$
we have from (\ref{jo1}) and (\ref{jo2})
\begin{eqnarray}
\left|\;u_{m,\:m}^{\varepsilon} \:+\:
\varepsilon \:u_{m,\:m}^{\varepsilon^2}\;
\right|
\; \leq \;
{\cal G}_{m}\:
\left(
1\:+\:I_{m,\:m}
\right)
\label{jo3_3}
\end{eqnarray}
Here
\begin{eqnarray}
{\cal H}_{m}\;=\;
m^2\:
\left(
C_4\:+\:
\varepsilon\, 8\, m\,
C_3  \left(
1\:+\:|\vec{y}|^3
\right)
\right)
\label{jo3_4}
\end{eqnarray}
\begin{eqnarray}
{\cal G}_{m}\;=\;
2 \,m \,C_3\:
\left(
|\vec{y}|\:+\:
\varepsilon \,2\,m
\left(
1\:+\:|\vec{y}|^2
\right)
\right)
\label{jo3_5}
\end{eqnarray}
Let $\bar{\cal H}_{m}$ and $\bar{\cal G}_{m}$ be some
positive constants.
Consider the function
\begin{eqnarray}
v_m \;=\;
\left(
1\:+\:\tilde{I}_{m,\:m}
\right)
\exp\left(
-\varepsilon^2 
\frac{\bar{\cal H}_{m}}{1 - \varepsilon \bar{\cal G}_{m}}\: t
\right)
\nonumber
\end{eqnarray}
for which after some straightforward calculations
we have
\begin{eqnarray}
L\: v_m \;\leq\;
\varepsilon^2 
\left(
\left(
{\cal H}_{m}\:-\:
\bar{\cal H}_{m}
\right)\:+\:
\varepsilon
\frac{\bar{\cal H}_{m}}{1 - \varepsilon \bar{\cal G}_{m}}
\left(
{\cal G}_{m}\:-\:
\bar{\cal C}_{m}
\right)
\right)
\cdot
\nonumber
\end{eqnarray}
\begin{eqnarray}
\cdot
\left(
1\:+\:I_{m,\:m}
\right)
\exp\left(
-\varepsilon^2 
\frac{\bar{\cal H}_{m}}{1 - \varepsilon \bar{\cal G}_{m}} \:t
\right)
\label{mmest}
\end{eqnarray}
Choosing now
\begin{eqnarray}
\bar{\cal H}_{m}\;=\;
m^2\:
\left(
C_4\:+\:
\varepsilon\,8\, m\,
C_3  \left(
1\:+\:c^3(\varepsilon)
\right)
\right)
\nonumber
\end{eqnarray}
\begin{eqnarray}
\bar{\cal G}_{m}\;=\;
2\, m\, C_3\:
\left(
c(\varepsilon)\:+\:
\varepsilon\,2\, m
\left(
1\:+\:c^2(\varepsilon)
\right)
\right)
\nonumber
\end{eqnarray}
and assuming that $\:\varepsilon\:$ is small enough
to guarantee $\:1 - \varepsilon \bar{\cal G}_{m}\:>\:0\:$
we get from (\ref{mmest}) that $\;L\:v_m\:\leq\:0\;$ on
the set $\:|\vec{y}|\:\leq\:c(\varepsilon)\:$ and hence
\begin{eqnarray}
\left<
v_m
\left(s_{t}^{\varepsilon}\right)
\right> 
\;\leq\;
\left<
v_m\left(0\right)
\right> 
\label{prop34}
\end{eqnarray}
Using that with probability one
\begin{eqnarray}
1\:+\:\tilde{I}_{m,\:m}
\left(s_{t}^{\varepsilon}\right)
\;\leq\;
\left(
1 + \varepsilon \bar{\cal G}_{m}
\right)
\left(
1\:+\:I_{m,\:m}
\left(s_{t}^{\varepsilon}\right)
\right)
\nonumber
\end{eqnarray}
\begin{eqnarray}
1\:+\:\tilde{I}_{m,\:m}
\left(s_{t}^{\varepsilon}\right)
\;\geq\;
\left(
1 - \varepsilon \bar{\cal G}_{m}
\right)
\left(
1\:+\:I_{m,\:m}
\left(s_{t}^{\varepsilon}\right)
\right)
\nonumber
\end{eqnarray}
and hence with probability one 
\begin{eqnarray}
\left({1 - \varepsilon \bar{\cal G}_{m}}\right) \;
\left(1\:+\:I_{m,\:m}
\left(s_{t}^{\varepsilon}\right)\right)
\exp\left(-
\varepsilon^2 
\frac{\bar{\cal H}_{m}}{1 - \varepsilon \bar{\cal G}_{m}} \:t
\right)
\;\leq\;
v_m\left(s_{t}^{\varepsilon}\right)
\nonumber
\end{eqnarray}
\begin{eqnarray}
v_m\left(0\right)
\;\leq\;
\left({1 + \varepsilon \bar{\cal G}_{m}} \right)\;
\left(1\:+\:I_{m,\:m}
\left(0\right)\right)
\nonumber
\end{eqnarray}
we have from (\ref{prop34}) the estimate
\begin{eqnarray}
\left<1\:+\:I_{m,\:m}
\left(s_{t}^{\varepsilon}\right)\right>
\;\leq\;
\frac{1 + \varepsilon \bar{\cal G}_{m}}
{1 - \varepsilon \bar{\cal G}_{m}} \;
\left<1\:+\:I_{m,\:m}
\left(0\right)\right> \;
\exp\left(
\varepsilon^2 
\frac{\bar{\cal H}_{m}}{1 - \varepsilon \bar{\cal G}_{m}} \:t
\right)
\nonumber
\end{eqnarray}
which for the following is rewritten in the form
\begin{eqnarray}
\max_{0 \leq t \leq L / \varepsilon^2}
\hspace{0.3cm}
\left<1\:+\:I_{m,\:m}
\left(s_{t}^{\varepsilon}\right)\right>
\;\leq\;
\bar{\cal D}_m \:
\left<1\:+\:I_{m,\:m}
\left(0\right)\right>
\label{toctho}
\end{eqnarray}
where
\begin{eqnarray}
\bar{\cal D}_m \;=\;
\frac{1 + \varepsilon \bar{\cal G}_{m}}
{1 - \varepsilon \bar{\cal G}_{m}}
\exp\left(
\frac{L \bar{\cal H}_{m}}{1 - \varepsilon \bar{\cal G}_{m}}
\right),
\hspace*{0.7cm}
\lim_{\varepsilon \rightarrow 0} \bar{\cal D}_m \;=\;
\exp\left(m^2 L C_4\right)
\nonumber
\end{eqnarray}

{\bf 10.}
Denoting
$\;u_{m,\:k}\:=\:u_{m,\:k}^{\varepsilon} \:+\:
\varepsilon
u_{m,\:k}^{\varepsilon^2}\;$
and introducing the new remainder 
\begin{eqnarray}
\begin{array}{llllll}
\tilde{R}_{m,\:k}  = R_{m,\:k} &-
&\bar{c}_2(m,\:k) \: u_{m-2,\:k} &-& 
\bar{c}_2^{\;*}(k,\:m) \: u_{m,\:k-2} &-  \\
\\
& & 
\bar{c}_1(m,\:k) \: u_{m-1,\:k} & - &
\bar{c}_1^{\;*}(k,\:m) \: u_{m,\:k-1} &- \\
\\
& & 
\bar{c}_3(m,\:k) \: u_{m-2,\:k+1} & - &
\bar{c}_3^{\;*}(k,\:m) \: u_{m+1,\:k-2} &-  \\
\\
& & 
\bar{c}_4(m,\:k) \: u_{m-2,\:k+2} & - &
\bar{c}_4^{\;*}(k,\:m) \: u_{m+2,\:k-2} &-  \\
\\
& & 
\bar{c}_5(m,\:k) \: u_{m-1,\:k+1}  & - &
\bar{c}_5^{\;*}(k,\:m) \: u_{m+1,\:k-1} &- \\
\\
& & 
\bar{c}_6(m,\:k) \: u_{m-1,\:k-1}  & - &
\bar{c}_7(m,\:k) \: u_{m,\:k} 
\end{array}
\nonumber
\end{eqnarray}
which admits the estimate (as follows from 
(\ref{jo1})-(\ref{jo2}))
\begin{eqnarray}
\left|\;\tilde{R}_{m,\:k}\;\right|
\; \leq \;
C_5 (m + k)^3
(1+ \varepsilon (m+k)) 
\left(1 +
\left| \vec{y} \right|^3
\right)
\left(
1 + 
\left|\;I_{m,\:k}\;\right|
\right)
\label{jo3_p}
\end{eqnarray}
we get from (\ref{jj76})
\begin{eqnarray}
\begin{array}{llllll}
L \: \tilde{I}_{m,\:k} & =
&\varepsilon^2 \left(
\bar{c}_2(m,\:k) \: \tilde{I}_{m-2,\:k} \right.& + &
\bar{c}_2^{\;*}(k,\:m) \: \tilde{I}_{m,\:k-2} & + \\
\\
& & \hspace*{0.6cm}
\bar{c}_1(m,\:k) \: \tilde{I}_{m-1,\:k} & + &
\bar{c}_1^{\;*}(k,\:m) \: \tilde{I}_{m,\:k-1} & + \\
\\
& & \hspace*{0.6cm}
\bar{c}_3(m,\:k) \: \tilde{I}_{m-2,\:k+1} & + &
\bar{c}_3^{\;*}(k,\:m) \: \tilde{I}_{m+1,\:k-2} & + \\
\\
& & \hspace*{0.6cm}
\bar{c}_4(m,\:k) \: \tilde{I}_{m-2,\:k+2} & + &
\bar{c}_4^{\;*}(k,\:m) \: \tilde{I}_{m+2,\:k-2} & + \\
\\
& & \hspace*{0.6cm}
\bar{c}_5(m,\:k) \: \tilde{I}_{m-1,\:k+1}  & + &
\bar{c}_5^{\;*}(k,\:m) \: \tilde{I}_{m+1,\:k-1} & + \\
\\
& & \hspace*{0.6cm}
\bar{c}_6(m,\:k) \: \tilde{I}_{m-1,\:k-1}  & + &
\left. 
\bar{c}_7(m,\:k) \: \tilde{I}_{m,\:k}\right) &
+\\
\\
& & \varepsilon^3 \; \tilde{R}_{m,\:k}\:
\end{array}
\label{jj76_00}
\end{eqnarray}

Let now $N$ be as in the theorem B. Using notation
$\:\vec{\cal V}(\,*\,;\,N)\:$ we can rewrite (\ref{jj76_00})
in the form of the following system
\begin{eqnarray}
L\:\vec{\cal V}(\tilde{I}_{m,\:k} ;\,N)\;=\;
\varepsilon^2 \,\bar{\cal K}_N\:
\vec{\cal V}(\tilde{I}_{m,\:k} ;\,N)\:+\:
\varepsilon^3\,
\vec{\cal V}(\tilde{R}_{m,\:k} ;\,N)
\label{star_01}
\end{eqnarray}
where the matrix $\:\bar{\cal K}_N\:$ is the same
as in (\ref{srk}).

The matrix $\:\bar{\cal M}_N(\tau)\:$ is assumed to be
the fundamental matrix solution of (\ref{srk}). That means
that the matrix 
$\:\bar{\cal M}_N^{-1}(\varepsilon^2 t)\:$ satisfies
\begin{eqnarray}
\frac{d}{d t}
\bar{\cal M}_N^{-1}(\varepsilon^2 t)\;=\;
-\varepsilon^2\,
\bar{\cal M}_N^{-1}(\varepsilon^2 t)\:\bar{\cal K}_N,
\hspace{1.0cm}
\bar{\cal M}_N^{-1}(0)\;=\;I
\label{star_02}
\end{eqnarray}
Applying the operator $\:L\:$ to the vector
$\:\bar{\cal M}_N^{-1}(\varepsilon^2 t)\, 
\vec{\cal V}(\tilde{I}_{m,\:k}(t) ;\,N)\:$
and taking into account (\ref{star_01}), (\ref{star_02})
we obtain
\begin{eqnarray}
L\:
\left(
\bar{\cal M}_N^{-1}(\varepsilon^2 t)\, 
\vec{\cal V}(\tilde{I}_{m,\:k}(t) ;\,N)
\right)\;=\;
\varepsilon^3\,
\bar{\cal M}_N^{-1}(\varepsilon^2 t)\, 
\vec{\cal V}(\tilde{R}_{m,\:k}(t) ;\,N)
\label{star_03}
\end{eqnarray}

From (\ref{star_03}) and Dynkin's formula
(see, for example \cite{Kushner}) it follows
that
\begin{eqnarray}
\left\langle
\bar{\cal M}_N^{-1}(\varepsilon^2 s_t^{\varepsilon})\, 
\vec{\cal V}(\tilde{I}_{m,\:k}(s_t^{\varepsilon}) ;\,N)\;-\;
\vec{\cal V}(\tilde{I}_{m,\:k}(0) ;\,N)
\right\rangle\;=
\nonumber
\end{eqnarray}
\begin{eqnarray}
=\;\varepsilon^3 \,
\left\langle
\int\limits_0^{\;s_t^{\varepsilon}}
\bar{\cal M}_N^{-1}(\varepsilon^2 \tau)\, 
\vec{\cal V}(\tilde{R}_{m,\:k}(\tau) ;\,N)\,d\tau
\right\rangle
\nonumber
\end{eqnarray}
or, equivalently
\begin{eqnarray}
\left\langle
\bar{\cal M}_N^{-1}(\varepsilon^2 s_t^{\varepsilon})\, 
\vec{\cal V}(I_{m,\:k}(s_t^{\varepsilon}) ;\,N)\;-\;
\vec{\cal V}(I_{m,\:k}(0) ;\,N)
\right\rangle\;=
\nonumber
\end{eqnarray}
\begin{eqnarray}
=\;\varepsilon 
\cdot 
\left\langle
\vec{\cal V}(u_{m,\:k}(0) ;\,N)\;-\;
\bar{\cal M}_N^{-1}(\varepsilon^2 s_t^{\varepsilon})\, 
\vec{\cal V}(u_{m,\:k}(s_t^{\varepsilon}) ;\,N)
\right\rangle\;+
\nonumber
\end{eqnarray}
\begin{eqnarray}
+\;\varepsilon^3 \,
\left\langle
\int\limits_0^{\;\;s_t^{\varepsilon}}
\bar{\cal M}_N^{-1}(\varepsilon^2 \tau)\, 
\vec{\cal V}(\tilde{R}_{m,\:k}(\tau) ;\,N)\,d\tau
\right\rangle
\label{star_04}
\end{eqnarray}

From (\ref{star_04}) we obtain
\begin{eqnarray}
\max_{0 \leq t \leq L / \varepsilon^2}
\hspace{0.3cm}
\left|
\left\langle
\bar{\cal M}_N^{-1}(\varepsilon^2 s_t^{\varepsilon})\, 
\vec{\cal V}(I_{m,\:k}(s_t^{\varepsilon}) ;\,N)\;-\;
\vec{\cal V}(I_{m,\:k}(0) ;\,N)
\right\rangle
\right|\;\leq
\nonumber
\end{eqnarray}
\begin{eqnarray}
\leq\;
\varepsilon
\cdot
\max_{0 \leq \tau \leq L}
\left|
\bar{\cal M}_N^{-1}(\tau)
\right| 
\max_{0 \leq t \leq L / \varepsilon^2}
\left\langle
2
\left|
\vec{\cal V}(u_{m,\:k}(s_t^{\varepsilon}) ;\,N)
\right|
\,+\,L
\left|
\vec{\cal V}(\tilde{R}_{m,\:k}(s_t^{\varepsilon}) ;\,N)
\right|
\right\rangle
\label{star_05}
\end{eqnarray}
Let us define now $\:[m]\:$ as the smallest integer
which is bigger or equal to $\,m$. 
Using (\ref{jo1}), (\ref{jo2}), (\ref{jo3_p}) and simple
inequalities like
\begin{eqnarray}
\left|\,\vec y\,\right|\;\leq\;1\:+\:
\left|\,\vec y\,\right|^2
\nonumber
\end{eqnarray}
\begin{eqnarray}
1\:+\:\left|\,I_{m,\:k}\,\right|\;\leq\;
2\,\left(1\:+\:
I_{\left[\frac{N}{2}\right],\:\left[\frac{N}{2}\right]}
\right),
\hspace{0.7cm}
m\:+\:k\;\leq\;N
\nonumber
\end{eqnarray}
we can obtain 
\begin{eqnarray}
\max \left\{
\left|\vec{\cal V}(u_{m,\:k};\,N)\right|,\,
\left|\vec{\cal V}(\tilde{R}_{m,\:k};\,N)\right|
\right\}\;\leq
\nonumber
\end{eqnarray}
\begin{eqnarray}
\leq\;
C_6 N^5\left(1+\varepsilon N\right)
\left(1 +
\left| \vec{y} \right|^3
\right)
\left(
1 + 
I_{\left[\frac{N}{2}\right],\:\left[\frac{N}{2}\right]}
\right)
\label{star_06}
\end{eqnarray}
Taking into account (\ref{toctho}) we have from (\ref{star_06})
\begin{eqnarray}
\max_{0 \leq t \leq L / \varepsilon^2}
\left\langle
2
\left|
\vec{\cal V}(u_{m,\:k}(s_t^{\varepsilon}) ;\,N)
\right|
\,+\,L
\left|
\vec{\cal V}(\tilde{R}_{m,\:k}(s_t^{\varepsilon}) ;\,N)
\right|
\right\rangle\;\leq
\nonumber
\end{eqnarray}
\begin{eqnarray}
\leq\;
C_6 N^5\left(1+\varepsilon N\right)
\left(1 +
c^3(\varepsilon)
\right)
(2+L)\bar{\cal D}_{\left[\frac{N}{2}\right]}
\left(
1 + 
I_{\left[\frac{N}{2}\right],\:\left[\frac{N}{2}\right]}(0)
\right)
\nonumber
\end{eqnarray}
which together with
\begin{eqnarray}
\max_{0 \leq \tau \leq L}
\left|
\bar{\cal M}_N^{-1}(\tau)
\right| 
\;\leq\;\exp(C_7 N^2)
\nonumber
\end{eqnarray}
and (\ref{star_05}) gives the final estimate we need
\begin{eqnarray}
\max_{0 \leq t \leq L / \varepsilon^2}
\hspace{0.3cm}
\left|
\left\langle
\bar{\cal M}_N^{-1}(\varepsilon^2 s_t^{\varepsilon})\, 
\vec{\cal V}(I_{m,\:k}(s_t^{\varepsilon}) ;\,N)\;-\;
\vec{\cal V}(I_{m,\:k}(0) ;\,N)
\right\rangle
\right|\;\leq
\nonumber
\end{eqnarray}
\begin{eqnarray}
\leq\;
\varepsilon\,
\left(1 +
c^3(\varepsilon)
\right)\, P\,
\left(
1 + 
I_{\left[\frac{N}{2}\right],\:\left[\frac{N}{2}\right]}(0)
\right)
\label{star_0f}
\end{eqnarray}
where
\begin{eqnarray}
P\;=\;
C_6 N^5\left(1+\varepsilon N\right)
(2+L)\bar{\cal D}_{\left[\frac{N}{2}\right]}
\exp(C_7 N^2)
\nonumber
\end{eqnarray}

Taking the limit $\;\varepsilon\:\rightarrow\:0\;$
we have from (\ref{star_0f}) the proof of the
theorem B with speed of convergence  
$\;\varepsilon \,c^3(\varepsilon)$.

{\bf 11.} To prove the remark to the theorem B
we apply the operator $\:L\:$ to the function
$\:\vec{a}_N(\varepsilon^2 t) \cdot 
\vec{\cal V}(\tilde{I}_{m,\:k}(t) ;\,N)\:$
with the result that 
\begin{eqnarray}
L\:
\left(
\vec{a}_N(\varepsilon^2 t) \cdot 
\vec{\cal V}(\tilde{I}_{m,\:k}(t) ;\,N)
\right)\;=\;
\varepsilon^3\,
\vec{a}_N(\varepsilon^2 t) \cdot 
\vec{\cal V}(\tilde{R}_{m,\:k}(t) ;\,N)
\label{vecfr}
\end{eqnarray}
The rest of the proof is just repeating  all the steps
from the previous point with the usage (\ref{vecfr}) instead of
(\ref{star_03}).

\subsection{Sketch of the Proof of the Theorem D}

\hspace{0.5cm}
{\bf 1.} The solution of the system (\ref{m02wn})
is a Markovian diffusion process in the $2$-dimensional
Euclidean space. Let $\:L\:$ be a generating differential operator 
of this stochastic process. Separating the orders according to
$\:\varepsilon\:$ we can represent $\:L\:$ in the form
\begin{eqnarray}
L \;=\; L_0 \:+\: 
\varepsilon^2 \: L_{\varepsilon^2}
\label{m03_0}
\end{eqnarray}
where the differential operators $\:L_0\:$ and
$\:L_{\varepsilon^2}\:$ are defined as follows
\begin{eqnarray}
L_0 \;=\; 
\frac{\partial}{\partial t} \; + \; 
\omega_0 \: 
\left( 
z \: \frac{\partial}{\partial x} - 
x \: \frac{\partial}{\partial z}
\right)
\nonumber
\end{eqnarray}

\begin{eqnarray}
L_{\varepsilon^2}  = 
\left(
\:z\:
\left(\Phi\:\vec{d}\cdot\vec{d} \:-\: \alpha\right)
\:+\:
x\:\Phi\:\vec{b}\cdot\vec{d}
\:-\:\Phi\:\vec{h}\cdot\vec{d}\:
\right)
\frac{\partial}{\partial z}\:+
\nonumber
\end{eqnarray}
\begin{eqnarray}
+\:\left(
\:\Phi\:\vec{h}\cdot\vec{h}\:-\:
2 x\:\Phi\:\vec{h}\cdot\vec{b}\:-\:
2 z\:\Phi\:\vec{h}\cdot\vec{d}\:+
\right.
\nonumber
\end{eqnarray}
\begin{eqnarray}
+\:\left.
x^2\:\Phi\:\vec{b}\cdot\vec{b}\:+\:
2 x z\:\Phi\:\vec{d}\cdot\vec{b}\:+\:
z^2\:\Phi\:\vec{d}\cdot\vec{d}\:
\right)
\frac{\partial^2}{\partial z^2}
\nonumber
\end{eqnarray}

{\bf 2.} Now we wish to calculate 
$\:L_{\varepsilon^2}\,I_{m,\:k}\,$.
Representing the operator $\:L_{\varepsilon^2}\:$ in the form
\begin{eqnarray}
L_{\varepsilon^2} \;=\; 
\left(\rule[0.25cm]{0cm}{0.25cm}
-\Phi \vec{h}\cdot\vec{d}\:+\:
\left[\:
\Phi \vec{b}\cdot\vec{d}\:-\: i\,\alpha \:+\:
i\,\Phi \vec{d}\cdot\vec{d}\:
\right]
\exp(i\:\omega_0\: t)\,I_{0,\: 1}
\:+
\right. 
\nonumber
\end{eqnarray}
\begin{eqnarray}
+\:\left.
\left[\:
\Phi \vec{b}\cdot\vec{d}\:+\: i\,\alpha \:-\:
i\,\Phi \vec{d}\cdot\vec{d}\:
\right]
\exp(-i\:\omega_0\: t)\,I_{1,\: 0}
\rule[0.25cm]{0cm}{0.25cm} \right)\cdot
\frac{\partial}{\partial z}\;+
\nonumber
\end{eqnarray}
\begin{eqnarray}
+\;
\left(\rule[0.25cm]{0cm}{0.25cm}
\Phi \vec{h}\cdot\vec{h}\:-\:2
\left[\:
\Phi \vec{h}\cdot\vec{b}\:+\: 
i\,\Phi \vec{h}\cdot\vec{d}\:
\right]
\exp(i\:\omega_0\: t)\,I_{0,\: 1}
\:-
\right. 
\nonumber
\end{eqnarray}
\begin{eqnarray}
-\;
2 \left[\:
\Phi \vec{h}\cdot\vec{b}\:-\: 
i\,\Phi \vec{h}\cdot\vec{d}\:
\right]
\exp(-i\:\omega_0\: t)\,I_{1,\: 0}
\:+\:
2 \left[\:
\Phi \vec{b}\cdot\vec{b}\:+\: 
\Phi \vec{b}\cdot\vec{d}\:
\right]\,I_{1,\: 1}\:+
\nonumber
\end{eqnarray}
\begin{eqnarray}
+\;
\left[\:
\Phi \vec{b}\cdot\vec{b}\:-\: 
\Phi \vec{d}\cdot\vec{d}\:+\: 
2\,i\,\Phi \vec{d}\cdot\vec{b}\:
\right]
\exp(i\:2\:\omega_0\: t)\,I_{0,\: 2}
\:+
\nonumber
\end{eqnarray}
\begin{eqnarray}
\left.
+\;
\left[\:
\Phi \vec{b}\cdot\vec{b}\:-\: 
\Phi \vec{d}\cdot\vec{d}\:-\: 
2\,i\,\Phi \vec{d}\cdot\vec{b}\:
\right]
\exp(-i\:2\:\omega_0\: t)\,I_{2,\: 0}
\rule[0.25cm]{0cm}{0.25cm} \right)\cdot
\frac{\partial^2}{\partial z^2}
\nonumber
\end{eqnarray}
taking into account (\ref{re4}), property {\bf b} 
and the expression
\begin{eqnarray}
\frac{\partial^2 I_{m, \: k}}{\partial z^2} \: = \:
\frac{m k}{2}\:I_{m-1,\: k-1}\:-
\nonumber
\end{eqnarray}
\begin{eqnarray}
-\:
\frac{m(m-1)}{4}\: \exp( i \:2\: \omega_0 \: t) \; I_{m-2,\: k} \; - \;
\frac{k(k-1)}{4}\: \exp(-i \:2\: \omega_0 \: t) \; I_{m, k-2} 
\nonumber
\end{eqnarray}
we obtain that $\:L_{\varepsilon^2}\, I_{m\:,k}\:$ is
given by the right hand side of (\ref{ccil}) with
$\:c_l(m,\,k)\:$ as follows
\begin{eqnarray}
\begin{array}{lll}
c_1(m, \:k) & = &
\hspace{0.3cm}
\frac{m}{2}\: 
\left[\:
\rule[0.2cm]{0cm}{0.2cm}
(m-2k-1)\,\Phi\, \vec{h}\cdot\vec{b}\:-\:
i\,(m+2k)\,\Phi \,\vec{h}\cdot\vec{d}\:
\right] \:
\exp(i\: \omega_0\: t)\\
\\
c_2(m, \:k) & = &
-
\frac{m(m-1)}{4} \:
\left[\:
\rule[0.2cm]{0cm}{0.2cm}
\Phi\, \vec{h}\cdot\vec{h}\:
\right] \: \exp(i\: 2\:\omega_0\: t)\\
\\
c_3(m, \:k) & = &
\hspace{0.3cm}
\frac{m(m-1)}{2} \:
\left[
\rule[0.2cm]{0cm}{0.2cm}
\:
\Phi\, \vec{h}\cdot\left(\vec{b}\:-\:i\,\vec{d}\:\right)\,
\right] \:
\exp(i\: 3\:\omega_0\: t)\\
\\
c_4(m, \:k) & = &
-
\frac{m(m-1)}{4}\:
\left[
\rule[0.2cm]{0cm}{0.2cm}
\:\Phi\,
\left(\vec{b}\:+\:i\,\vec{d}\:\right)\,
\cdot
\left(\vec{b}\:-\:i\,\vec{d}\:\right)\,
\:\right]
\:\exp(i\: 4\:\omega_0\: t)\\
\\
c_5(m, \:k) & = &
\hspace{0.3cm}
\frac{m}{2} 
\left[
\rule[0.2cm]{0cm}{0.2cm}
\alpha \;-\;
(m+k)\,
\Phi\, \vec{d}\cdot\vec{d}\:+\:
(k-m+1)\,
\Phi\, \vec{b}\cdot\vec{b}
\right.\:+\\
\\
& &
\hspace{0.3cm}
\left.
i\,(2k+1)\,
\Phi\, \vec{b}\cdot\vec{d}\:
\rule[0.2cm]{0cm}{0.2cm}
\right] \:
\exp(i\: 2\:\omega_0\: t)\\
\\
c_6(m, \:k) & = &
\hspace{0.3cm}
\frac{m k}{2}\:
\Phi\, \vec{h}\cdot\vec{h}\\
\\
c_7(m, \:k) & = &
-\frac{m+k}{2}\:\alpha\:+\:
\frac{4mk-m(m-1)-k(k-1)}{4}
\:\Phi\, \vec{b}\cdot\vec{b}\:+\\
\\
& &
\hspace{0.3cm}
\frac{4mk+m(m+1)+k(k+1)}{4}
\:\Phi\, \vec{d}\cdot\vec{d}\:+\:
i\,\frac{m^2-k^2}{2}\:
\:\Phi\, \vec{d}\cdot\vec{b}
\end{array}
\nonumber
\end{eqnarray}

{\bf 3.}
Now, like in the proof of  theorem B,
we want to show that there exist 
for fixed $m$ and $k$
functions 
$\:g_l(m,\: k)\:$ 
bounded in $t$
and
satisfying
\begin{eqnarray} 
\frac{\partial g_l(m,\: k)}{\partial t}\;+\;
c_l(m,\: k)\;=\;\breve{c}_l(m,\: k),
\hspace*{1.0cm} l\;=\;1,\ldots,7
\nonumber
\end{eqnarray}
We shall show it for $\:l\:=\:4\:$ and the rest can be
done by analogy.

Substituting in the expression for $\:c_4(m,\:k)\:$
the Fourier series of the vectors $\:\vec{b}\:$ and
$\:\vec{d}\:$ we obtain
\begin{eqnarray} 
c_4(m, \:k) \; = \;
-\frac{m(m-1)}{2} \: 
\sum \limits_{p,\:l\;=\;-\infty}^{\infty}
\left[\Phi\:
\left(\vec{b}_p\,+\,\vec{d}_p\right) 
\cdot 
\left(\vec{b}_l\,-\,\vec{d}_l\right) 
\right] 
\: \exp(i\,\mu_{p,\:l} t) \;=
\nonumber
\end{eqnarray}
\begin{eqnarray} 
=\;\breve{c}_4(m,\:k)\:-\:
\frac{m(m-1)}{4}
\sum \limits_{\mu_{p,\:l}\:\neq\:0}
\left[\Phi\:
\left(\vec{b}_p\,+\,\vec{d}_p\right) 
\cdot 
\left(\vec{b}_l\,-\,\vec{d}_l\right) 
\right] 
\: \exp(i\,\mu_{p,\:l} t) 
\nonumber
\end{eqnarray}
where $\:\mu_{p,\:l}\:=\:\nu_p-\nu_l+4\omega_0\,$.

So the problem will be solved if the series
\begin{eqnarray} 
c_4(m, \:k) \; = \;i\,
\frac{m(m-1)}{4}
\sum \limits_{\mu_{p,\:l}\:\neq\:0}
\frac{
\Phi\:
\left(\vec{b}_p\,+\,\vec{d}_p\right) 
\cdot 
\left(\vec{b}_l\,-\,\vec{d}_l\right) 
}{\mu_{p,\:l}} 
\: \exp(i\,\mu_{p,\:l} t) 
\nonumber
\end{eqnarray}
converges and can be differentiated term by term.
The absolute convergence and differentiability can be easily shown
using (\ref{cond1}) and (\ref{cond2}) with the final estimates
\begin{eqnarray}
\left|\:
g_4(m,\:k)
\:\right|
\;\leq\; 
\frac{m(m-1)}{2}
\frac{\left|\Phi\right|}{\delta_f^2} 
\left(
\sum \limits_{p = -\infty}^{\infty}
\left|\vec{b}_{p}\right|\:+\:
\left|\vec{d}_{p}\right|
\right)^2
\nonumber
\end{eqnarray}
\begin{eqnarray}
\left|\:
\frac{\partial g_4(m,\:k)}{\partial t}
\:\right|
\;\leq\; 
\frac{m(m-1)}{2}
\,\left|\Phi\right|\,
\left(
\sum \limits_{p = -\infty}^{\infty}
\left|\vec{b}_{p}\right|\:+\:
\left|\vec{d}_{p}\right|
\right)^2
\nonumber
\end{eqnarray}

{\bf 4.} The rest follows the simplified version of the
proof of the theorem B.

\end{document}